%% file: 0-main.tex
\documentclass[onecolumn,compsoc]{IEEEtran}
\usepackage{verbatim} 
\usepackage{algorithm}
\usepackage{algpseudocode}
\usepackage{eurosym}
\usepackage{float}
\usepackage{amssymb}
\usepackage{amsmath}
\usepackage{graphicx}
\usepackage[T1]{fontenc}
\usepackage[utf8]{inputenc}
\usepackage{flushend}
\usepackage{chngcntr}
\usepackage{url}
\usepackage{xspace}
\usepackage{listings}
\usepackage{fancyhdr}
\usepackage[normalem]{ulem}
\usepackage{soul}
\usepackage{todonotes}
\usepackage{enumerate}
\usepackage[shortlabels]{enumitem}
\usepackage[noadjust]{cite}
\usepackage[switch]{lineno}
\usepackage{multirow}
\usepackage{xcolor}

\counterwithin{paragraph}{subsection}

\newcommand\copyrighttext{%
	\footnotesize \textit{\textbf{Authors' manuscript.}}\quad Published in IEEE Access 2020. The final publication is available at IEEE via \texttt{http://dx.doi.org/10.1109/ACCESS.2020.3032239}. }
\newcommand\copyrightnotice{%
	\begin{tikzpicture}[remember picture,overlay]
		\node[fill=lightgray,draw] at (8.5,2.2) {\parbox{0.58\textwidth}{\copyrighttext}};
	\end{tikzpicture}%
}

\begin{document}

\title{Using Fault Injection to Assess Blockchain Systems in Presence of Faulty Smart Contracts}

\author{\
\IEEEauthorblockN{
Ákos Hajdu\IEEEauthorrefmark{2}, 
Naghmeh Ivaki\IEEEauthorrefmark{1}, 
Imre Kocsis\IEEEauthorrefmark{2}, 
Attila Klenik\IEEEauthorrefmark{2}, 
László	Gönczy\IEEEauthorrefmark{2},
Nuno Laranjeiro\IEEEauthorrefmark{1},
Henrique Madeira\IEEEauthorrefmark{1},
András Pataricza\IEEEauthorrefmark{2}
}

\IEEEauthorblockA
{
	\IEEEauthorrefmark{1}CISUC, Department of Informatics Engineering\\University of Coimbra, Portugal\\
    \{naghmeh,cnl,henrique\}@dei.uc.pt
}
\\
    \IEEEauthorblockA{\IEEEauthorrefmark{2}Department of Measurement and Information Systems\\Budapest University of Technology and Economics, Hungary\\
	\{hajdua,ikocsis,klenik,gonczy,pataric\}@mit.bme.hu}
}

\maketitle

\copyrightnotice

\input{0-abstract.tex}


\input{1-intro.tex}

\input{2-relatedwork.tex}
\input{4-FI-SC.tex}
\input{5-experimental-setup.tex}
\input{6-results}
\input{8-conclusion}

\section*{Acknowledgments}
This work has been supported by the project ``Advanced Analytics for Empirical Assessment of Cloud Resilience'', granted by the bi-lateral FCT-NKFIH program Portugal-Hungary; the European Union's Horizon 2020 research and innovation program under the Marie Sklodowska-Curie grant agreement No 823788 ``ADVANCE'', the BME-Artificial Intelligence TKP2020/IK grant of NRDI; the NRDI Fund based on the charter of bolster issued by the NRDI Office under the auspices of the Ministry for Innovation and Technology; and the {\'U}NKP-19-3 New National Excellence Program of the Ministry for Innovation and Technology.

\bibliographystyle{IEEEtran}
\bibliography{references}

\end{document}

%% file: 0-abstract.tex

\begin{abstract}
Blockchain has become particularly popular due to its promise to support business-critical services in very different domains (e.g., retail, supply chains, healthcare). Blockchain systems rely on complex middleware, like Ethereum or Hyperledger Fabric, that allow running smart contracts, which specify business logic in cooperative applications. The presence of software defects or faults in these contracts has notably been the cause of failures, including severe security problems.
In this paper, we use a software implemented fault injection (SWIFI) technique  to assess the behavior of permissioned blockchain systems in the presence of faulty smart contracts. We emulate the occurrence of general software faults (e.g., missing variable initialization) and also blockchain-specific software faults (e.g., missing \emph{require} statement on transaction sender) in smart contracts code to observe the impact on the overall system dependability (i.e., reliability and integrity).
We also study the effectiveness of formal verification (i.e., done by solc-verify) and runtime protections (e.g., using the \emph{assert} statement) mechanisms in detection of injected faults. Results indicate that formal verification as well as additional runtime protections have to complement built-in platform checks to guarantee the proper dependability of blockchain systems and applications.
The work presented in this paper
allows smart contract developers to become aware of possible faults in smart contracts and to understand the impact of their presence.
It also provides valuable information for middleware developers to improve the behavior (e.g., overall fault tolerance) of their systems.
\end{abstract}

%% file: 1-intro.tex
\section{Introduction}
\label{sec:intro}
Blockchain is a software-based distributed ledger technology, which, just like in a hard copy ledger, is a way for cooperating partners to store and track transaction records  \cite{stallings_internet_2017}. Its growing popularity is due to core properties as decentralization, immutability, security and transparency \cite{puthal2018blockchain} and to the fact that it naturally applies to many contexts, from supply chain management through financial markets to healthcare \cite{odair_networked_2017, CurberaBL, iansiti_truth_2017}.

In a blockchain environment, code supplied by some of the users of the system defines transaction logic that is application-specific. The Ethereum technology \cite{wood2014ethereum} made such programs, similar to stored procedures and named \emph{smart contracts} \cite{szabo1996smart}, popular as a general mechanism for executing client-proposed ledger transactions. Cryptographically linked (\emph{chained}) \emph{blocks} of the proposed transactions have to pass a group consensus among a set of nodes (\emph{peers}), maintaining the distributed, shared ledger.
The group consensus accounts for the order as well as any consequences of smart contract based transactions, which lead to modifications in the ledger state.

Blockchain systems are commonly viewed as being extremely secure and dependable, but the reality is much more nuanced. Regarding security, while consensus may probably be attack-resistant, it does not address the fact that deployed smart contract code may hold software defects, including vulnerabilities \cite{li2017survey, atzei2017survey}, allowing for certain executions to harm the system. Regarding the general concept of dependability, key properties of the system like ledger integrity, service reliability and service availability are a function of a complex interplay between different mechanisms that target different non-functional properties, which are difficult to check using traditional verification techniques.

Within such a complex environment, quality (i.e., the absence of software faults) of smart contracts plays a vital role. Residual bugs in smart contracts (software faults that may be activated due to absent or inadequate protection mechanisms, or that were not detected by verification activities like testing or static analysis) have been linked to failures in blockchain systems and are known to be the cause of severe security problems \cite{nist2019,atzei2017survey}, as, for instance, the infamous DAO attack demonstrated \cite{dhillon2017dao}.

In this paper, \textbf{we present a software implemented fault injection (SWIFI) technique that has been tailored for evaluating the behavior of blockchain systems in the presence of faulty smart contracts}. We target smart contracts written in the popular Solidity language \cite{ethereum_solidity_2019}, which is being supported by an increasing number of blockchain platforms. Starting with a general software fault model for software fault injection as basis \cite{duraes2006emulation}, we eliminate rare software faults and faults that do not apply to our context (e.g., faults that cannot affect programs written in the Solidity language) and complement the fault model with \textbf{smart contract-specific faults}, based on 515 faults listed at the NIST National Vulnerability Database \cite{nist2019} and blockchain and smart contract bugs presented in the literature. Based on the resulting fault model, we then inject software faults into the smart contract abstract syntax tree to generate faulty smart contracts which are then executed on a blockchain platform.

The blockchain system is observed and analyzed considering the following perspectives: i) \textit{reliability (or correctness)}: divergence from correct behavior regarding the external behavior of the system (observable by the client); ii) \textit{integrity evaluation}: the ledger state integrity violation (either observable by the client or not).
We complement the empirical assessment using \textit{formal verification}, i.e., static analysis for possible detection of injected faults and verification of incorrect functioning of smart contracts (to understand which faults and incorrect behavior would be captured before deployment and which would likely escape to production).

We carried out an experimental evaluation using \textit{Hyperledger Fabric} \cite{androulaki2018hyperledger}, the market-leading business blockchain platform, and a set of 15 smart contracts (5 base versions, 5 versions modified to hold extensive protections against invalid inputs, and 5 versions holding no protection) in which we injected faults, which resulted in a total of 651 faulty versions. Results show the distinct effects of the different types of injected faults and strongly suggest that using formal verification and additional code-level protection mechanisms can prevent the occurrence or activation of different types of faults, thus being of utmost importance for developing and deploying critical blockchain applications.

The main contributions of this paper are:

\begin{itemize}
    \item A \textit{smart contract software fault model} of 51 fault types (33 general and 18 smart contract specific);
    \item The definition of an \textit{approach for injecting faults} into smart contracts, which is independent from the programming language used;
    \item A \textit{fault injector tool} for Solidity smart contracts;
    \item Experimental evaluation that shows i) the impact of different faults on the behavior of a blockchain system; and ii) the effectiveness and the complementary nature of \emph{smart contract formal verification}, \emph{runtime platform checks}, and \emph{contract-level protections} under our fault model .
\end{itemize}

The paper is organized as follows. The next section, Section \ref{sec:relatedwork}, presents background and related work and
Section \ref{sec:FM} presents the smart contract fault model and the fault injection approach used. Section \ref{sec:experimental_setup} presents the experimental setup and implementation and the evaluation process followed. Section \ref{sec:results} presents and discusses the results obtained and Section \ref{sec:threats} highlights the threats to the validity of the work. Finally, Section \ref{sec:conclusion} concludes the paper.

%% file: 2-relatedwork.tex
\section{Background and Related Work}
\label{sec:relatedwork}

In this section, we present the necessary background on blockchain technology and smart contracts and review existing works that address dependability (and security) in this environment.

\subsection{Blockchain Technologies}
Blockchain is, in practice, a digital implementation of the hardcopy transaction ledger \cite{zheng_blockchain_2018}, which acts as a highly secure and resilient database and execution environment for the business logic deployed on it in the form of smart contracts. Nowadays, digital ledgers are used in businesses and organizations in a variety of applications, such as movement of assets or properties, recording contracts, buy-sell deals, and liabilities documentation. 
A large number of increasingly mature blockchain platforms are available today, facilitating the creation of public access, ``unpermissioned'' consensus participation networks with cryptocurrencies as well as the ones that can be accessed only by specific business consortia in a fully permissioned manner. 

Ethereum is a key technology in both worlds. A core part of its specification is a simple bytecode virtual machine for smart contract execution, the \emph{Ethereum Virtual Machine} (EVM) \cite{wood2014ethereum}. The EVM is a general-purpose, Turing-complete virtual machine that provides access to the ledger state largely via key-value style CRUD (create, read, update and delete) operations, and Solidity is the most popular language targeting the EVM \cite{ethereum_solidity_2019}. 
However, Solidity is also becoming available on blockchain platforms other than Ethereum. For instance, the Hyperledger Burrow \cite{Hyperledger_Burrow} EVM implementation can be deployed in Hyperledger Fabric.

Solidity is a statically typed, contract-oriented language that has high-level facilities such as complex data types, multiple inheritance and libraries, with a syntax resembling ECMAScript. The Solidity programming model natively handles cryptocurrency (Ether) movements between pseudonymous (cryptographic key authenticated) parties; and EVM-based execution is defined in terms of the calling party offering up-front cryptocurrency payment as fuel (\emph{gas}) for smart contract function call executions. If the offered gas runs out before termination, the call fails. Regardless, these peculiarities can be easily hidden or circumvented in business-oriented blockchain networks and applications. Our paper does not require a more in-depth introduction of either Ethereum or the EVM; the interested reader is kindly referred to the above references.

\begin{figure}[t]
\centering
\includegraphics[width=0.8\linewidth]{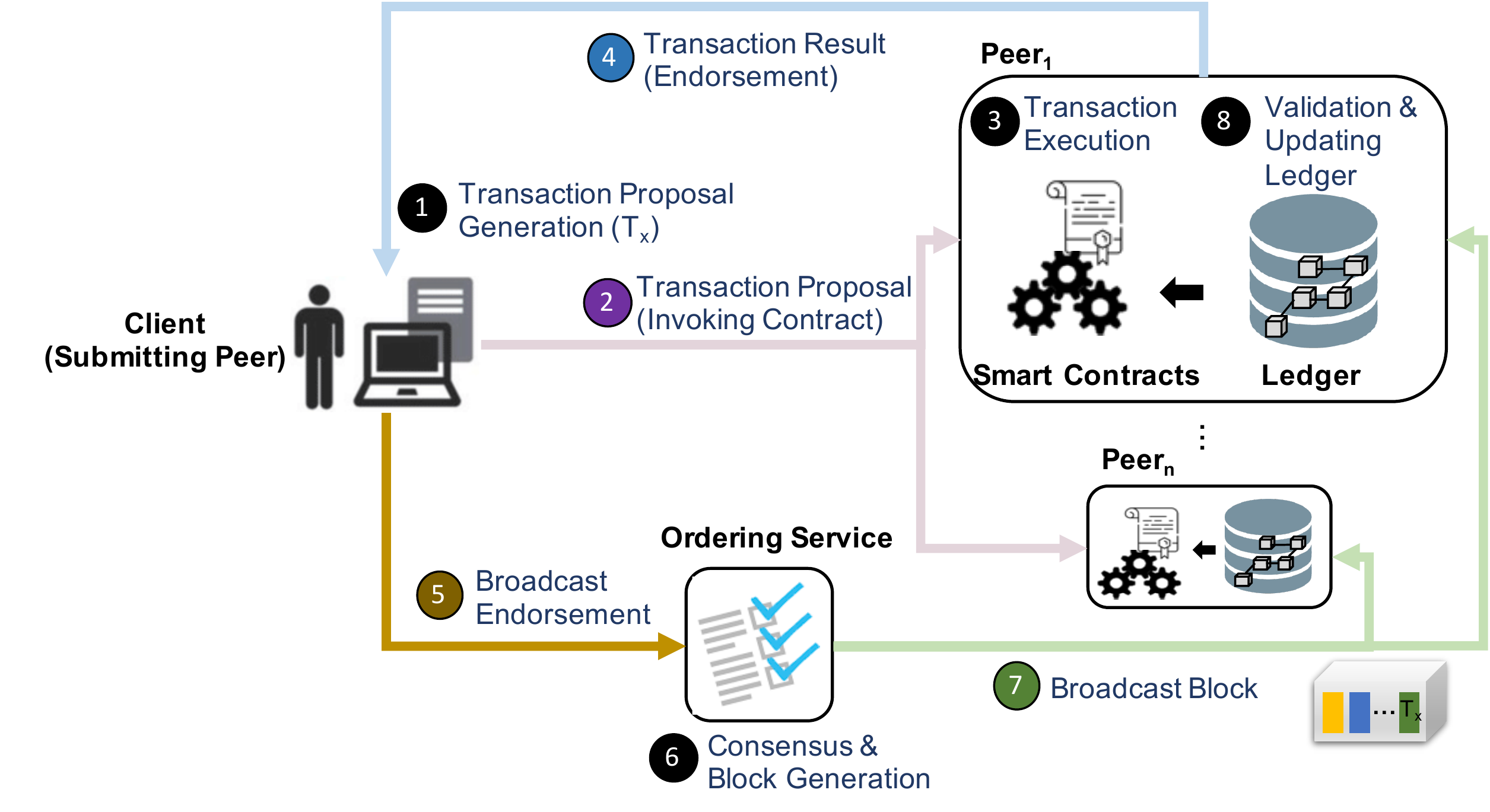}
\caption{Transaction execution in Hyperledger Fabric.}
\label{fig:blockchainarchitecture}
\end{figure}

For our experiments, we chose Hyperledger Fabric \cite{androulaki2018hyperledger}, the platform most completely supporting the creation of ``bespoke'', business consortium specific, permissioned blockchain systems. In Hyperledger Fabric, a set of \emph{organizations} as a \emph{consortium} form a blockchain network from peers controlled by them. The network is directly accessible only to their systems and users. 

Compared to most blockchain platforms, transaction execution and block consensus is fairly unique in Hyperledger Fabric. Fig. \ref{fig:blockchainarchitecture} presents an overview. The core idea is that a client asks peers of organizations to \emph{endorse} a proposed transaction (essentially, smart contract function call) by executing it on their current ledger state and replying with the so-called \emph{read-write set} upon success -- instead of modifying their ledger state. The read-write set contains the (versioned) set of variables and the updates a successful execution would perform. Given enough matching (signed) endorsements, the \emph{ordering service} accepts the proposal and includes it in the next block, to be distributed to the peers for ledger state updates. The full protocol, which is described in detail in \cite{androulaki2018hyperledger} and in the Hyperledger Fabric documentation, is summarized below and presented in Fig. \ref{fig:blockchainarchitecture}: 

\begin{enumerate}
\item A client assembles and digitally signs a transaction proposal;
\item The client sends the transaction proposal directly to one or more so-called \emph{endorsing} peers of one or more organizations participating in the blockchain;
\item After checking the identity of the client, the peers execute the transaction (smart contract function call) using their current ledger state -- but do not update it; in essence, performing \emph{transaction simulation};
\item The peers respond to the client with a digitally signed \emph{read-write set} of such ledger variables that the actual transaction execution would read and write;
\item The client collects the responses and submits them to an ordering service (provided by a disinterested third party, or a group consensus protocol across the organizations);
\item The ordering service performs a consensus protocol and forms blocks of the multiparty-endorsed transaction proposals;
\item The ordering service disseminates the blocks to the peers of the organizations;
\item Peers validate the transactions in the blocks and modify their ledger state. 
\end{enumerate}

\subsection{Related Work}

In business or mission-critical environments, the presence of software bugs in a blockchain system is a problem that can bring severe consequences to the business or mission itself, impairing its dependability \cite{atzei2017survey}. This is aggravated by the fact that, in some cases, the programming languages used for writing the contracts are not mainstream languages (e.g., Solidity) and developers may lack the necessary expertise, leading to the deployment of contracts holding residual bugs. 

Despite several attempts in the literature to define best practices for implementing secure and fault-free smart contracts \cite{delmolino2016step, wohrer2018smart},  developing contracts without faults seems to be very difficult, if possible. As smart contracts often manipulate (or provide accounting for) valuable assets, verification of smart contracts 
is playing an increasingly important role \cite{zheng_blockchain_2018}, but as the blockchain technology is still relatively recent, there are just a few, and not really effective, tools~\cite{atzei2017survey,harz2018safer,miller2018smart} (e.g., static analyzers~\cite{bhargavan2016formal,grishchenko2018semantic,tsankov2018securify,luu2016making,nikolic2018finding,mythril,slither}, theorem provers~\cite{hildenbrandt2017kevm,hirai2017defining,sergey2018scilla,BeckertHerdaKirstenEA2018,nehai2019deductive}, SMT-based tools~\cite{leonardo2018smt,solcverify,lahiri2018formal,permenev2019verx}).

Testing tools~\cite{truffle,soliditycoverage} have been developed to detect both bugs and vulnerabilities, as well as formal approaches to assure the correctness of smart contracts~\cite{atzei2017survey,harz2018safer,miller2018smart}.

Most of these approaches are adaptations of, or built on, existing methods, inheriting their advantages and limitations. For example, static and symbolic analyzers are effective at finding vulnerable patterns in smart contracts~\cite{bhargavan2016formal,grishchenko2018semantic,tsankov2018securify,luu2016making,nikolic2018finding,mythril,slither}, but might yield a high rate of false alarms and miss bugs related to the business logic. Formal  methods  provide  a  sound  basis,  but usually require formal specification~\cite{BeckertHerdaKirstenEA2018,solcverify,permenev2019verx} or assisted proofs~\cite{hildenbrandt2017kevm,hirai2017defining,sergey2018scilla,BeckertHerdaKirstenEA2018,nehai2019deductive}.

Our work is focused on using the software-implemented fault injection (SWIFI) technique, which is a vital tool for evaluating the dependability of software systems, especially critical systems. Typical SWIFI approaches include the injection of code changes in certain components of the system (that emulate the most frequent types of bugs introduced during development) or the direct injection of the effects of faults (e.g., faulty data or user inputs). Regardless of the technique, the goal is to be able to evaluate how a faulty component may affect the behavior of the overall system, which also allows for understanding the effectiveness of the fault tolerance mechanisms in place \cite{natella_assessing_2016}. 

To the best of our knowledge, no previous work has proposed software fault injection for studying the dependability and security of smart contracts.
Fuzzing tools~\cite{Jiang2018,ponte2018fuzzing,he2019learning} are related to our work in the sense they also generate inputs to check for vulnerabilities. However, existing fuzzers only consider the original (existing) contracts and do not include fault injection to evaluate the impact of faults in smart contracts.

Our approach of using software fault injection can be considered as a complementary technique to assess the effects of both common programming errors and smart contract-specific faults. Furthermore, software fault injection can also be used as a benchmark to evaluate the precision and recall of existing tools such as static analyzers and fuzzers.

%% file: 4-FI-SC.tex
\section{Fault Injection into Smart Contracts}
\label{sec:FM}

This section describes the overall approach for how reliable  a blockchain system is in the presence of faulty smart contracts. The approach relies on the execution of smart contract transactions accompanied with the formal verification of possible defects, which works a complement to the runtime execution (to understand if certain defects could be caught before runtime). In the next subsections we present: i) an overview of the approach; ii) the definition of the fault model used and its application to smart contracts; iii) the workload generation process; iv) the process of formal verification applied to contracts and v) how the collected data is analyzed.

\subsection{Approach Overview}

\begin{figure*}[h]
\centering
  \includegraphics[width=0.80\linewidth]{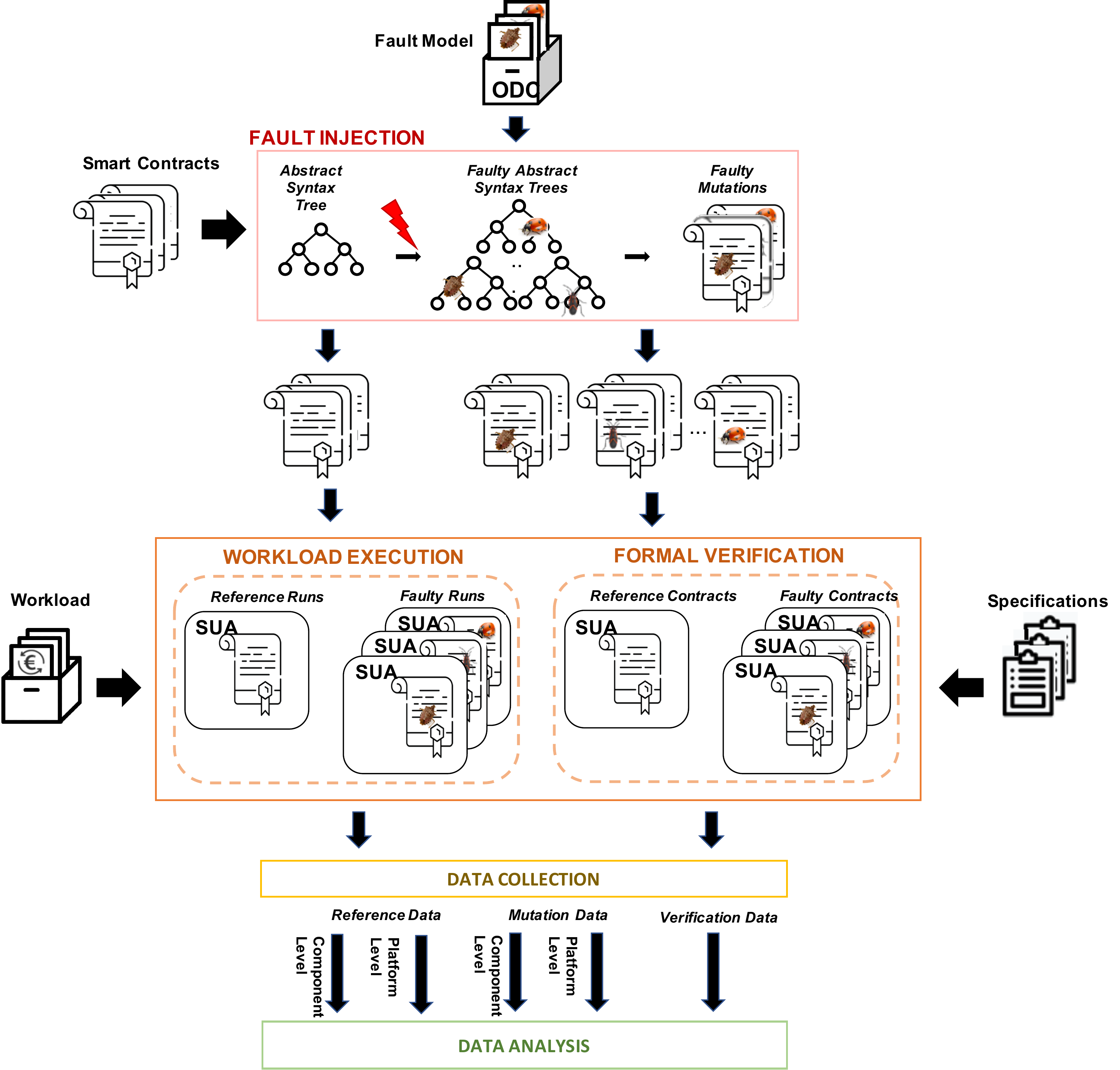}
  \caption{Approach overview.}
  \label{fig:process}
\end{figure*}

Fig.\ref{fig:process} presents a general view of the fault injection environment, the involved components, and the process followed to inject faults into the system under assessment (SUA). 

In general, to apply a fault injection technique, we need to address the following main questions:  which parts of the system are targeted for fault injection (i.e., fault injection location), what faults to be injected, and when to inject the faults? 
In this work, the target of fault injection within the blockchain system is the \textbf{smart contract}. Thus, it is one of the inputs for the approach (depicted at the top left-hand side of Fig. \ref{fig:process}). One of the challenges involved in the use of a fault injection technique is the definition of a \textbf{fault model} (at the top in Fig. \ref{fig:process}) that should be composed of representative types of faults, which will be injected, in our case, into the smart contracts. The definition of the fault model essentially covers general cases of typical programming mistakes \cite{chillarege1992orthogonal, duraes2006emulation} and specific cases of smart contract related faults \cite{nist2019}.
  
To \textbf{inject faults into smart contracts}, we integrate changes (i.e., fault models similar to mutations but based on realistic fault assumptions) into the source code of the smart contracts. The implementation of these changes (i.e., faults) strongly depends on the programming language used to write the smart contract. To eliminate this dependency our \textbf{fault injector} first generates an Abstract Syntax Tree (AST), which is an abstract representation of the syntactic structure of source code, for a certain smart contract source code provided as input. We then inject individual faults into the AST, which results in several faulty ASTs, each with exactly one injected fault. Then the faulty ASTs are converted back to code (Solidity code, in our case), resulting in faulty versions smart contracts. 

Both the \textbf{fault-free and faulty smart contracts are then executed} individually on the blockchain system, in an isolated environment. Fault-free runs are used as a reference to evaluate the results of faulty runs. A challenge at this point is the generation of \textbf{workloads} that are representative and allow activating the artificially introduced faults during the execution of transactions. 

We complement the assessment by performing \textbf{formal verification} over both reference contracts and the respective mutated contracts that are annotated with \textbf{specifications} (e.g., contract invariants, pre- and post-conditions). This allows to assess the effects of \emph{protection mechanisms} in smart contract code (e.g., integrity-protecting error detecting assertions and observability-increasing return statements) and to understand which faults and incorrect behavior would be captured before deployment. During the assessment, the system under assessment (SUA) is monitored during the execution of the fault-free and faulty smart contracts and all necessary \textbf{data are collected} and stored in log files, for later analysis.

After finishing the runs, the \textbf{analysis of the data} is performed from the reliability (i.e., correct external and internal behavior) and integrity (i.e., ledger state integrity) perspectives at two levels: i) \emph{component level}, which refers to the direct analysis of the behavior of the container running the smart contracts, allowing us to understand the impact of the faults on the container in which the smart contract is being executed; ii) \emph{platform level}, in which the analysis is performed at level of the whole blockchain platform, i.e., to understand the impact of faults on the whole blockchain system setting. At each of these analysis levels, we have reference data corresponding to the outcome of fault-free runs (i.e., component level reference data and platform level reference data) and mutation data corresponding to the outcome of faulty smart contract runs (component level mutation data and platform level mutation data).

The details regarding the definition of fault model, formal verification, and criteria used for behavior assessment are presented in the following sections. 

\subsection{Fault Model Definition}
\label{sec:fault-model-definition}

Our fault model uses a combination of software faults covering two cases: i) general software faults \cite{chillarege1992orthogonal} and ii) smart contract specific faults, which include 515 faults listed at the NIST National Vulnerability Database  \cite{nist2019} and also blockchain and smart contract bugs presented in the literature \cite{destefanis2018smart,luu2016making,grishchenko2018semantic,tsankov2018securify, wan2017bug}.

To build the fault model, we use the information provided by the Orthogonal  Defect  Classification (ODC) \cite{chillarege1992orthogonal}, which provides a useful foundation for emulating software faults. ODC is usually used as a source of information to provide insights and feedback regarding the quality of software and the development process, but it also provides a useful foundation regarding the emulation of software faults. Building on ODC, an extensive classification of representative software faults was carried out in \cite{duraes2006emulation}, in which the identification and classification of software faults are done from a fault injection perspective so that the identified faults can be easily emulated in code. 

According to \cite{duraes2006emulation}, the most representative types of defects, i.e., of software faults are (in order): i) Assignment (e.g., the value assigned to a variable is incorrect); ii) Checking (e.g., the input data is not validated correctly ); iii) Interface (e.g., error in function call); iv) Algorithm (e.g., incorrect implementation of an algorithm); and v) Function (e.g., a functionality is affected). The nature of a defect type (i.e., the ODC qualifier \cite{chillarege1992orthogonal}) can be one of the following cases: Missing construct (e.g., missing function call), Wrong construct (e.g., the wrong variable used in parameter of a function call), and Extraneous construct (e.g., extraneous function call). 

We started the definition of our fault model based on the general results in \cite{duraes2006emulation}, which, in total, identifies $62$ software faults. From this list, we removed the ones that do not apply to our context, which uses a specific programming language (Solidity) and respective compiler. For instance, \textit{Missing case: statement(s) inside a switch construct (MCS)} and \textit{Wrong branch construct - goto instead break (WBCI)} are not applicable because there are no \texttt{goto} and \texttt{switch} statements in Solidity, or \textit{Missing parameter in function call (MPFC)} and \textit{wrong data types or conversion used (WSUT)} are also not applicable because they are caught by the compiler. After this analysis, $8$ faults out of the $62$ were excluded from the fault model.

We also excluded the less frequent faults (those appearing with a frequency lower than $1\%$ (Refer to \cite{duraes2006emulation} for more details about frequency of the faults). \textit{Wrong parenthesis in logical expression in parameters of function call (WPLP)} is one example of these faults.
This resulted in a total of $21$ faults being excluded. For the time being, we have a list of $33$ frequent generic software fault types, identified in the last column of Table~\ref{tab:fm} with the letter \textbf{G}, that apply to our field study. 

In a second step, with the goal of increasing the representativeness of the fault model, we completed the list of faults with smart contract/Solidity specific faults. These faults are identified in the last column of Table \ref{tab:fm} with the letter \textbf{S}. To define the specific faults, we collected and analyzed all smart contract faults reported in the National Vulnerability Database (NVD) \cite{nist2019} at the time of writing (a total of $515$ vulnerabilities), as well as the blockchain and smart contract bugs presented and analyzed in the literature~\cite{destefanis2018smart,luu2016making,grishchenko2018semantic,tsankov2018securify, wan2017bug}. Table \ref{tab:vuln}, summarizes all vulnerabilities reported in NVD, categorized into $8$ distinct types and from which we focus on the more frequent ones.

\begin{table}[h]
    \centering
    \caption{Smart contract specific faults reported in NVD \cite{nist2019}.}
    \label{tab:vuln}
    \begin{tabular}{c|c|c}
         \begin{tabular}{@{}c@{}} \textbf{Type of Vulnerability} \end{tabular} &
         \begin{tabular}{@{}c@{}}\textbf{Count} \end{tabular} &
         \begin{tabular}{@{}c@{}}\textbf{Percentage} \end{tabular} \\
         \hline
         \hline
         \textcolor{red}{Integer Overflow or Wraparound} & \textcolor{red}{474}  & \textcolor{red}{92\%} \\
         \hline
         \textcolor{olive}{Improper Access Control} & \textcolor{olive}{18}  & \textcolor{olive}{3,5\%} \\
         \hline
         \textcolor{cyan}{Input Validation} & \textcolor{cyan}{9} & \textcolor{cyan}{1,7\%} \\
         \hline
         Use of Cryptographically Weak Pseudo-Random Number Generator (PRNG) & 8 & 1,6\% \\
         \hline
         Use of Insufficiently Random Values & 2 & 0,4\%  \\
         \hline
         Integer Underflow (Wrap or Wraparound) & 2 & 0,4\% \\
         \hline
         Integer Buffer Errors & 1 & 0,2\% \\
         \hline
         Out-of-bounds Read & 1 & 0,2\% \\
         \hline
    \end{tabular}
\end{table}

As shown in Table \ref{tab:vuln}, the most frequent vulnerability is \textit{Integer Overflow or Wraparound} ($92\%$ of all reported vulnerabilities), which occurs when the result of an arithmetic computation exceeds the range of a type. When this occurs, the value may wrap to become a negative number or a very small number. For example, $127 + 1$ becomes $-128$,  instead of $128$, when one signed byte ($8$ bits) is used to store the result. Overflows do not cause a run-time error and can silently lead to incorrect calculations and undefined behavior. This becomes a severe issue when the result of an arithmetic computation is used to control loops, determine the size in operations such as memory allocation, or take some other action decision (e.g., used to control an authentication or authorization process).
In some systems, it may even impact performance. Smart contract developers can avoid integer overflows using recognized integer handling libraries, such as \textit{SafeMath} \cite{dourlens_safemath_2017}. Thus, missing calls to these safe integer handling functions may lead to integer overflows. For this reason, fault $25.a$ (Missing calls to \textit{SafeMath} (MCSM)) is defined under this category of faults (i.e., Integer Overflow or Wraparound) as a special case of the ODC fault \texttt{Missing function call (MFC)}. 

\begin{table}[t]
\centering
  \caption{Fault model.}
  \includegraphics[width=0.62\linewidth]{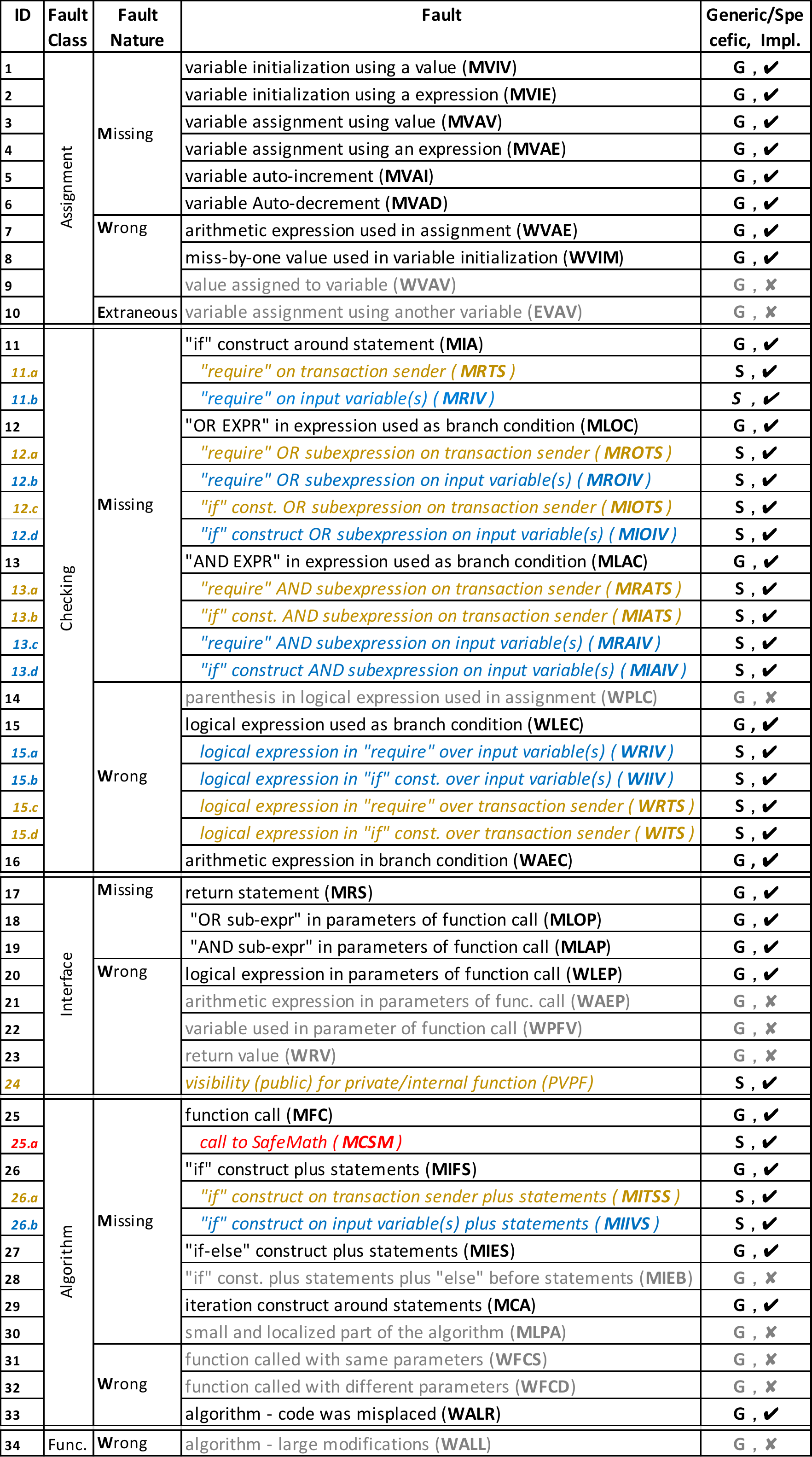}
  \label{tab:fm}
\end{table}

The next frequent vulnerability, although not as frequent as the Integer Overflow or Wraparound, is \textit{Improper Access Control} ($3.5\%$ of all reported vulnerabilities). This occurs, when a smart contract does not restrict or incorrectly restricts access to a certain resource or function from an unauthorized user. Several faults in smart contracts may introduce such access control weaknesses, in particular, those that include incorrect or missing conditions required to be verified for the sender of the transactions or incorrect visibility defined for the resources.
Faults 
$11.a$ (Missing \textit{require} on transaction sender (MRTS)), 
$12.a$ (Missing \textit{require} OR subexpression on transaction sender (MROTS)), 
$12.c$ (Missing \textit{if} construct OR subexpression on transaction sender (MIOTS)), 
$13.a$ (Missing \textit{require} AND subexpression on transaction sender (MRATS)), 
$13.b$ (Missing \textit{if} construct AND subexpression on transaction sender (MIATS)), 
$15.c$ (Wrong logical expression in \textit{require} over transaction sender (WRTS)), 
$15.d$ (Wrong logical expression in \textit{if} construct over transaction sender (WITS)), 
$24$ (Wrong visibility (public) for private/internal function (PVPF)), and 
$26.a$ (Missing \textit{if} construct on transaction sender plus statements (MITSS)) are defined under this category of faults (i.e., Improper Access Control) as special cases of the corresponding ODC faults.

\textit{Input Validation} is the next frequent smart contract vulnerability ($1.7\%$ of all reported vulnerabilities), which refers to wrong or missing validation of inputs (input values are not validated correctly or are not validated at all), 
something that usually takes place at the beginning of transactions to revert the whole transaction when a condition does not hold. 
Wrong inputs may affect data flow and control flow of the transaction leading to various problems, such as robustness or security issues. The possible input validation checks in smart contracts are very diverse, depending on the type of input data and the business logic expressed by the contracts, but to emulate this kind of vulnerability, such validation checks must be eliminated or modified (e.g., by altering the checking conditions). Faults 
$11.b$ (Missing \textit{require} on input variable(s) (MRIV)), 
$12.b$ (Missing \textit{require} OR subexpression on input variable(s) (MROIV)), 
$12.d$ (Missing \textit{if} construct OR subexpression on input variable(s) (MIOIV)), 
$13.c$ (Missing \textit{require} AND subexpression on input variable(s) (MRAIV)), 
$13.d$ (Missing \textit{if} construct AND subexpression on input variable(s) (MIAIV)), 
$15.a$ (Wrong logical expression in \textit{require} over input variable(s) (WRIV)), 
$15.b$ (Wrong logical expression in \textit{if} construct over input variable(s) (WIIV)), and 
$26.b$ Missing \textit{if} construct on input variable(s) plus statements (MIIVS)) are defined under this category of vulnerabilities as special cases of the corresponding ODC faults.

Finally, we find \textit{Use of Cryptographically Weak Pseudo-Random Number Generator (PRNG)} ($1.6\%$ of all reported vulnerabilities), which occurs when a non-cryptographic PRNG or a PRNG's algorithm that is not cryptographically strong is used in a context where security is important. For instance, if a non-cryptographic PRNG algorithm is used to generate a session ID (i.e., used for authentication and authorization), then an attacker might be able to easily guess the session ID and gain access to restricted resources. Although being an important security issue, since the impact of this vulnerability is very tightly coupled to the intention of the attacker and to the logic involved in the transactions and smart contracts, for the time being, we do not include it in our fault model.

As shown in the last column of Table \ref{tab:fm}, we implemented all smart contract specific faults and most (in total covering 78.4\%) of the software generic faults (identified with \checkmark). We left a few of the generic faults unimplemented (the ones that appear in light grey in the table and identified with \textit{X}), due to the fact that they do not seem to be very common and relevant in smart contracts according to the literature \cite{destefanis2018smart,luu2016making,grishchenko2018semantic,tsankov2018securify, wan2017bug}. 

\subsection{Workload Generation}
\label{subsec:test-set-generation}
Workload generation, in our context, is largely equivalent to automated test case generation. We are not aware of any such mature tool, so as part of our research, we defined a workload generation process that generates values for smart contract function call input parameters based on based on their \emph{type}, \emph{literals} appearing in the code, and \emph{randomly}, as follows:

\begin{itemize}[-]
    \item \textit{Type-based inputs}: generating input values for each parameter based on its type. For Booleans, we generate both true and false, while for integers, we generate the minimum, the maximum and zero. Input values for reference types, such as strings and arrays, are generated recursively for different lengths, including the length of zero.
    
    \item \textit{Literal-based inputs}: Inputs are often compared to literals, which determines the control-flow inside the function. Therefore, we also extract each literal from the function and use them as inputs to the parameters with matching type. Furthermore, for integer literals $l$ we also include $l+1$ and $l-1$ to catch off-by-one errors.

    \item \textit{Random inputs}: generating a configurable number of random values for certain types. For integers, we generate random values within the minimum and maximum range. We also generate random strings and arrays by generating random elements of their sub-type and shuffling them randomly for different lengths.

\end{itemize}

\subsection{Formal Verification}
We complement the empirical assessment with using \emph{formal verification} (i.e., static code analysis), in order to understand how difficult it would be to detect injected faults and incorrect functioning of smart contracts. To do so, we execute a formal verification technique, namely \emph{modular program verification} \cite{muller2002modular} over reference contracts and the respective mutated contracts. 
In modular program verification, the smart contracts source code should be annotated with \emph{specifications} (e.g., contract invariants, pre- and postconditions) of the expected behavior. Each annotated contract is translated into a formal representation in an intermediate verification language~\cite{barnett2006boogie}, having mathematically precise semantics. Each function in this representation is then translated into a set of formulas (so-called verification conditions) that establish the correctness of the function.

The verifier discharges these formulas automatically using SMT (Satisfiability Modulo Theories) solvers \cite{barrett2018satisfiability} by checking their validity. The modular verification approach used in this work targets the \emph{functional correctness} of contracts, with respect to different kinds of failures, including expected and unexpected failures. An \emph{expected failure} is a failure that is caused due to an explicit guard defined by developers (e.g., Solidity \texttt{require}, \texttt{revert} statements). An unexpected failure is any other failure (e.g., \texttt{assert}, overflow). A contract is reported as \emph{correct} if there is no unexpected failure and all of its transactions that are finished without an expected failure, satisfy their specification.

\subsection{Behavior Assessment}
\label{sec:behavior-assessment}
The last step of the approach is related to the evaluation of the behavior of the blockchain system in the presence of the injected faults, which is carried out in perspective with the fault-free runs. In this work, we concern about the faults' effects on the system's \textbf{reliability} (in \textit{lato sensu}) and \textbf{integrity}.
 Notice that integrity errors may, or may not, affect the reliable behavior of the blockchain system.

\textit{Reliability evaluation} mostly concerns divergence from correct behavior (i.e., regarding the external, observable, behavior). Evaluation of the \textit{external behavior} is relatively simple in the case of smart contracts. We can check whether the expected outcome of the transaction (success/failure) and the return value(s) (if any) match the fault-free runs. 

Regarding \textit{integrity evaluation}, the key aspect involved is to verify the ledger integrity, which can be done by comparing the read/write sets of each transaction executed by a faulty contract with that of the reference contract. Integrity is maintained only if all key-value pairs in the he read/write sets are exactly the same.
In this work, the integrity verification is done using the transaction read/write set feature of Hyperledger Fabric \cite{androulaki2018hyperledger}, in a way that it neither interferes with workload execution nor requires additional instrumentation and potential change coverage checking (as getter functions would). Although such changes could be detected to some extent by getter functions on the contract, the read/write sets allow us to catch parasite side effects as well. 

%% file: 5-experimental-setup.tex
\section{Experimental Evaluation}
\label{sec:experimental_setup}

This section presents the implementation of the approach presented previously. It explains how target smart contracts are selected and prepared for testing, how fault injection is implemented, how workload is generated and executed, how formal verification is executed, and how the test and formal verification results are analyzed.

\subsection{Smart Contracts Selection and Preparation}

To the best of our knowledge, neither a definitive general taxonomy for smart contract functionality, nor representative smart contract sets for security assessment, formal safety/liveness analysis or dependability analysis exists yet. The existing business use cases and token taxonomies, while useful, do not provide meaningful source code template libraries. Thus, we started by selecting five commonly-used contracts (i.e, representative in this case), also with the intention of covering a large cross-section of smart contract use cases. For each of the five \textit{base contracts}, we actually generate an unprotected version and a fully protected version of it, leading to a total of 15 smart contracts, in a process described in the next paragraphs.
The five \textit{base contracts} used as basis for our experiments are the following:

\begin{itemize}
 \item \textbf{State Machine:} The state machine contract \cite{stateMachineContract} realizes a supply-chain use case, where a single product is tracked across multiple parties. A temperature and humidity sensor constantly reports readings about the conditions of the product. The contract tracks whether the transport conditions satisfy predefined compliance criteria. This contract is representative for the \textit{permissioned state machine/business process} of the distributed ledger world. It is typically deployed on permissioned and private blockchains ~\cite{lahiri2018formal}.

\item \textbf{Wallet:} The wallet contract \cite{Contracts} realises a simple Ether store, where users of the contract can deposit and withdraw Ether. It is representative of \textit{bank-like ``cryptocurrency wallet''} and refundable escrow functionalities from the cryptocurrency domain. It is typically deployed on unpermissioned and public blockchains. It is worth noting that its implementations include the reentrancy bug that is cause of the DAO hack ~\cite{dhillon2017dao}.
    
\item \textbf{Token:} The token contract \cite{Contracts} implements the important aspects of an ERC20 standard-compliant fungible token \cite{anonerc20}, which has a limited supply and can be transferred between users, either individually or in batches. It is representative of a \textit{very general pattern} found across smart contracts \cite{bec2018cve}.
    
\item \textbf{Refund Escrow:} The escrow contract \cite{escrowContract} makes it possible for an escrower (the contract owner) to manage deposits incoming from multiple sources for designated payees. The contract also provides operational states like enabling/disabling refunding, and withdrawal for a beneficiary user. It is \textit{typically deployed on unpermissioned, public blockchains}.
    
\item \textbf{Storage:} The storage contract \cite{Contracts} provides a \textit{permissioned key-value storage} for users to store information encoded as integers, which is associated with their accounts. The contract has a designated administrator who has access to every storage slot regardless of ownership. 
    
\end{itemize}

As mentioned, for each of the above 5 \textit{base contracts}, we actually have the following different versions (described in further detail in the next paragraphs).

\begin{itemize}
    \item \textbf{Base contract:} The original unmodified contract, just as described previously;
    \item \textbf{Stripped contract:} A modified version of the original contract, where protection mechanisms have been explicitly removed for evaluation purposes.
    \item \textbf{Protected contract:} A modified version of the original contract, where protection mechanisms have been added.
\end{itemize}

Each of the \textit{base smart contracts} contain a ``reasonable'' level of defensive mechanisms (e.g., using \texttt{require} statements for error detection purposes), which stem from the business purpose of the contract (e.g., for input checking or access control), under the assumption that the implementation and their execution is fault-free. Solidity, as a language, provides assertion-style error detection mechanisms, and EVM error detection covers some basic situations as, e.g., array indexing out of bounds. For handling the detected errors, Solidity uses state-reverting exceptions and EVM safely reverts all changes made to the state. In general, exceptions can emanate from the following situations: i) Invalid operations in-contract, such as misindexing arrays; ii) Errors with calling/creating other contracts and certain ledger operations; iii) Failing application-level checks using the \texttt{assert}, \texttt{require} and \texttt{revert} \cite{ethereum_solidity_2019}.

The \textit{stripped smart contracts} were created to allow us understanding the impact of faults on contracts that lack protection mechanisms. If an unprotected contract may be used to harm the overall system, it is important to understand the potential impact of the presence of a certain software fault. The stripped smart contracts were created by simply stripping out the \texttt{require} statements from the base contracts.

Finally, we generated fully \textit{protected smart contracts}, mostly as a way of understanding if such mechanisms could be sufficient to protect the contracts from certain types of faults.
The protected smart contracts were created through the following two types of modifications. First, we include \texttt{assert} statements into each function as post-conditions (i.e., checked directly before the return of the ``protected'' functions). The assertion predicates were created manually, after carefully considering the valid and invalid function-local and contract states after execution.It is worth noting that none of the five selected contracts include \texttt{assert} statements. Second, where applicable, we extend each function to return part of its ledger side effect as a return value. Originally, most functions of the five selected contracts terminates without returning any values. The intention of these two modifications is to obtain a superior contract (in terms of error detection/mitigation) as, on one hand, failed assertions suppress integrity-compromising state changes (at the expense of accessibility - failing calls do not modify the ledger); on the other hand,  return values allow blockchain clients to perform error/failure detection themselves. In blockchain systems like Hyperledger Fabric, a client can choose to abandon a transaction \textit{after} execution, but still \textit{before} commitment. Note this latter case (i.e., error/failure detection by clients) may be only theoretically feasible,
as, in practice, factors ranging from constrained resources to the need to use gateway-based blockchain access can prevent clients from cross-checking expected and actual ledger modifications.

Given the above descriptions, we call a base smart contract, its stripped version, and its protected version a \textbf{contract family}.

\subsection{Fault Injection Implementation}
The faults defined in the fault model, presented earlier in Section \ref{sec:fault-model-definition}, are injected into all versions of the selected smart contracts resulting in multiple faulty versions of each original smart contract. A faulty contract will carry exactly one single artificially introduced fault. Smart contracts written in Solidity have to be compiled to bytecode using a compiler tool like Solc~\cite{solc}. Solc generates an \emph{abstract syntax tree} (AST) from the solidity source code, which can be easily manipulated to emulate faults. We have implemented fault injection as a series of transformations over the generated abstract syntax tree of the smart contracts, which is extracted from Solc~\cite{solc}.

Each fault is defined by a \emph{condition} and an \emph{action}. We read the AST, and whenever the specified condition holds for a given AST node, we transform it using the action. For example, to inject a Missing \textit{if} construct around statement (MIA) fault, the condition checks whether the current node being analysed is an \textit{if} statement. If the condition holds, the action replaces the node with its children (i.e., the body of the \emph{if} statement).

Sometimes there are multiple candidates, within the same smart contract, that can be used for injecting a certain fault (i.e., multiple nodes matching the condition). Whenever this occurs, we generate one faulty contract for each of the matching locations, resulting in several mutated contracts (again, we emphasize that each contract will hold a single fault). After injecting the faults, we then serialize the ASTs back to Solidity source code, so that it can be compiled to bytecode and deployed on the blockchain.

\subsection{Workload Generation for the Selected Contracts}
\label{subsec:test-seq}

During our experiments we evaluated the feasibility of using automatically generated call sequences (i.e., a random workload). On one hand, due to certain typical characteristics of Solidity smart contracts (e.g., only a few addresses actually having test-transferable tokens in deployed and test-initialized smart contract) an automatic test case generator tends to create an unacceptably high ratio of calls that fail by specification. Thus, in this context, the high number of failing ``garbage'' transactions (due to their randomness) would shadow the interesting failures occurring in regular user-contract interactions. On the other hand, certain software faults can be only triggered by very specific input sequences. For these reasons, we manually defined a  workload (i.e., a sequence of calls), exercising the target functions to observe the potential effects of the injected faults in every contract mutation. The construction of the workload was based on the combination of basic \emph{black-box} and \emph{white-box} testing principles. 

Regarding the \textbf{\textit{black-box perspective}}, we used the specification and the interface, to uncover the following aspects of each contract: i) the \emph{actors} interacting with the contract; ii) the \emph{classification of functions} based on whether they have side-effects on the ledger state or not, i.e., transaction and query functions, respectively; and iii) the \emph{input parameter space} of individual functions. Based on these aspects of the contracts, and from a \textit{black-box perspective}, the workload is generated as follows: 

\begin{itemize}[-]
    \item Issue transaction function calls by mixing the actors and parameter value space (detailed in the following paragraphs).
    \item Issue the whole set of available query function calls. This mimics user-initiated self-check queries.
    \item For both transaction and query function calls, use multiple actor identities, starting with the actors who should not be allowed to perform the given action. 
    \item When a function has input parameters, use \emph{equivalence partitioning} and \emph{interval testing} principles to exercise the function with meaningful inputs. Equivalence partitioning typically splits the domain of an input parameter into valid and invalid partitions (derived from the business logic), allowing testers to pick a single value from each partition as a representative input. Interval testing picks additional values from the partitions/ranges, usually from around their boundaries (if defined) to catch off-by-one errors. The combination of the two approaches reduces the number of required function calls for building a representative test sequence.
\end{itemize}

We also considered a \textit{\textbf{white-box perspective}}, whenever the contract exhibited a state-machine-like (or workflow-like) behavior.  We first construct the state transition graph of the contract by inspecting its implementation. The transitions in the graph drive the test sequence construction to achieve the highest possible state and transition coverage. Note that the complete coverage by a single test sequence is impossible if the state transition graph of the contract has multiple terminal strongly connected components (e.g., multiple sink states).

It is worth mentioning that, for the sake of readability, we do not differentiate between transaction and query function calls in the remainder of the paper. Note, however, that query function calls do not produce a read-write set and only the endorsement phase of the Fabric consensus applies to them.

\subsection{Test Harness and Execution}
Our experimental environment is a highly instrumented Hyperledger Fabric lab deployment, equipped with the Ethereum Virtual Machine (EVM) implementation of Hyperledger Burrow. The environment uses Hyperledger Caliper for most experiment automation tasks. We use the environment to perform \emph{test runs}: \emph{executing the contract family workload (test sequence) of a specific mutation}.
In our environment, the execution of a test run follows the below order:

\begin{enumerate}
    \item The system is (re)initialized which means that the Hyperledger nodes (endorsers and orderers) are set (or reset) in a clean initial state, with an empty blockchain and the Burrow EVM support installed. Then, the contract under assessment is deployed into the endorsing peers and the appropriate accounts are filled with local Ether to provide transferable cryptocurrency for certain contracts.
    
    \item For each transaction in a test sequence of a contract family:
    \begin{enumerate}
        \item The (Caliper) client sends the transaction request to each endorsing peer.
        \item The endorsing peers forward the call to the EVM implementation, running fenced-off in a Docker container.
        \item If the EVM execution terminates without error, the read-write sets are sent back to the client. Otherwise (upon EVM error, or if the Fabric peer terminates processing after the specified timeout) the client receives an error signal.
        \item If all endorsements permit (i.e., n-out-of-n consensus), the client sends them to the ordering service, which immediately creates a one-transaction block and distributes that to the peers for \emph{commitment} (inclusion in the current ledger state). It is worth noting that, while strictly sequential transaction execution with one-transaction blocks and n-out-of-n consensus are not production-typical for Hyperledger Fabric, for our purposes these are admissible and useful simplifications.
        \item The client proceeds with the next transaction only if the pending has been committed or can be deemed failed.
    \end{enumerate}
\end{enumerate}

During the test runs, detailed \emph{transaction-level} data is collected from multiple sources. Caliper/client-side observations provide the majority of transaction details, such as timing of life-cycle phases, side effects returned by the platform and general status information, for example, errors in case of unsuccessful endorsement. Low-level EVM runtime data is also collected from contract-side logs.

The collected data set is post-processed to match and compare the transactions of faulty contracts to their corresponding transaction pair in the reference contracts. The key comparison points are detailed in Section \ref{subsec:result-analysis}. 

\subsection{Formal Verification and Specifications}

For performing the formal verification, we execute a formal verification tool, namely solc-verify~\cite{solcverify}, over the smart contracts. Solc-verify is an automated modular verifier for Solidity smart contracts. It takes contract source code annotated with \emph{specifications} (e.g., contract invariants, pre- and post-conditions) and discharges verification conditions using modular program verification and SMT solvers. Solc-verify targets the \emph{functional correctness} of contracts and checks for possible assertion violations (Solidity \texttt{assert} statements) and overflows as implicit specification. Moreover, it has a feature for additional specifications such as:

\begin{itemize}[-]
    \item \emph{Contract invariants} that must hold after the constructor, and before and after every transaction;
    \item \emph{Function pre- and post-conditions} that are assumed before the function and checked after the function;
    \item \emph{Loop invariants} to specify loops with expressions that must hold before, after and in every iteration;
    \item \emph{Modification specifiers} for fine grained specification of the contract state that a function can modify.
\end{itemize}

In this work, for each contract family, a single set of specifications is created and used to annotate the base contracts, the stripped, and the protected contracts. The specifications are created manually for each base contract, using the facilities (invariants, preconditions, etc.) described in Section \ref{sec:behavior-assessment} and \cite{solcverify}. Table \ref{tab:contract-stats} presents key statistics about the number of assert and require/revert statements, and the formal specification statements for each contract family.

\begin{table}[h]
    \centering
    \caption{Contract statistics by types and families.}
    \label{tab:contract-stats}
    \begin{tabular}{c|c|c|c|c|c|c}
         \begin{tabular}{@{}c@{}}\textbf{Contract Family} \end{tabular} & 
         \begin{tabular}{@{}c@{}}\textbf{Test Length} \end{tabular} & 
         \textbf{Variant} & \textbf{Asserts} & \textbf{Requires} & 
         \begin{tabular}{@{}c@{}}\textbf{Formal Specifications} \end{tabular} 
         & \textbf{LOC} \\
         \hline\hline
         \begin{tabular}{@{}c@{}}State Machine \end{tabular} 
         & 252 & Base & - & 10 & 35 & 181 \\
         & & Stripped & - & - & 35 & 124 \\
         & & Protected & 7 & 10 & 35 & 198 \\
         \hline
         Wallet 
         & 19 & Base & - & 2 & 6 & 28 \\
         & & Stripped & - & - & 6 & 27 \\
         & & Protected & 3 & 2 & 7 & 39 \\
         \hline
         Escrow
         & 183 & Base & - & 9 & 31 & 243 \\
         & & Stripped & - & - & 31 & 219 \\
         & & Protected & 8 & 9 & 33 & 265 \\
         \hline
         Storage 
         & 72 & Base & - & 5 & 13 & 67 \\
         & & Stripped & - & - & 13 & 62 \\
         & & Protected & 8 & 5 & 13 & 83 \\
         \hline
         Token 
         & 151 & Base & - & 6 & 10 & 72 \\
         & & Stripped & - & - & 10 & 43 \\
         & & Protected & 6 & 6 & 10 & 92 \\
         \hline
    \end{tabular}
\end{table}

Solc-verify translates each smart contract into an intermediate verification language~\cite{barnett2006boogie} and performs modular verification~\cite{muller2002modular}, i.e., for each function it is checked whether its specification (e.g., invariants, post-conditions, assertions) hold, given the pre-conditions. The main feature that makes modular verification efficient is that when a function calls another one, the callee is replaced with its specification. However, this might introduce false alarms if the specification is not detailed enough. Note that, as solc-verify targets functional correctness (i.e., safety property and not liveness \cite{solcverify}), a contract that always reverts is considered correct (or safe) by the verifier.

\subsection{Verification and Protection Levels in Smart Contracts}
\label{subsec:protectionLevel}
Our experiments resemble, from a verification perspective, the life-cycle of a smart contract. Thus, some of the injected faults might be captured at different levels. In our experimental setup that is using Hyperledger Fabric running Solidity smart contracts using the Hyperledger Burrow EVM, this translates to the following 4 protection levels (i.e., including formal verification). Note that names in bold denote the terminology used in the figures presented in next Section.

\begin{itemize}[-]
    \item \textbf{Formal Verification}: applied before deployment (e.g., static analysis), it may detect a certain portion of existing faults \emph{before}.
    
    \item \textbf{Contract self-check}: specification-derived defensive programming constructs in the source/byte code (e.g., \texttt{require} statements in Solidity) and additional protective measures present in the source code (e.g., error detection by invariant checking directly before function return, using \texttt{assert} statements).
    
    \item \textbf{Runtime platform check}: error detection mechanisms in the EVM runtime implementation (e.g., detection of array out of bounds indexing, insufficient Ether balance for transfer requests, invalid address format, or runaway execution depleting gas) and error detection mechanisms in Hyperledger Fabric -- execution attempt timeout detection and result matching over the endorsing peer set. Both contract self-checks and runtime platform checks provide \emph{error detection and processing}; and in our case, limited fault tolerance through what is effectively a \emph{rollback} of the affected smart contract call.
    
    \item \textbf{Output invariant check}: a blockchain client may be also able to detect an execution error based on the return values it receives (assuming the availability of an appropriate test oracle). In general, it has, at least, failure detection capabilities (after a transaction has been committed to the ledger), given the necessary output-based observability and test oracle. Additionally, in a blockchain system where the client can also be involved in transaction consensus -- as is the case with Hyperledger Fabric -- the client can also choose to \emph{abort} the transaction attempt.
    
\end{itemize}

\subsection{Result Analysis}
\label{subsec:result-analysis}
As described above, each faulty contract is executed against its family-specific workload. Contract execution was performed exactly once for each test case; repeating the experiments showed a very high level of determinism (as expected, slight differences arise only in timing and ``runaway'' execution terminations).

The guiding principle of our assessment of the test case executions is that we aim at comparing behavior during executions on a faulty contract versus its corresponding fault-free reference.
For instance, the execution behavior of a faulty protected contract should be compared to the execution behavior of the fault-free protected contract, and for the same workload. Analysing runtime behavior of the faulty contracts without comparing it to the corresponding reference run would not be meaningful as some test sequence steps are intended to fail also in baseline and protected cases (e.g., authorization checking). 

In each execution, \textit{a transaction is deemed successful} if and only if i) all of its endorsements are successful and match and ii) it is successfully ordered and reported as committed by all endorsing peers. Additionally, verification was performed for each faulty contract, taking into account the supplied, contract family specific verification properties as well as the \texttt{require} and \texttt{assert} statements present. The resulting observations that we subject to data analysis include the following key characteristics and error/failure model for each faulty contract version:

\begin{itemize}
    \item \textbf{Observation identifier}: which includes \emph{contract family, version (i.e., base, stripped, protected), fault ID, mutation instance ID}.
    
    \item \textbf{Source code formal verification result}: which indicates whether the fault-free and faulty contracts were successfully distinguished by the formal verification tool. The result is reported as \emph{true positive}, \emph{true negative}, \emph{false positive}, or \emph{false negative}. Note that the \emph{true positive verification results} indicate the caught faulty mutations. Contracts execution results are important for the remaining cases (the false and true negatives).
    
    \item \textbf{Abort failure}: the result is a boolean (true or false) that indicates whether there was at least one transaction in the faulty contract that was aborted, while it did not fail in the reference case. \emph{Abort failure} shows that contract or platform checks detected at least one transaction invocation failure correctly. Specific error codes \cite{errorCodes}, which are not documented here, allow us to distinguish between the contract-level and platform-level checks' success or failure. Note that failing ``normal'' checks (e.g., \texttt{require} statements) can also be distinguished from the source code fault indicating error detecting ones. This failure is detectable by the client.
    
     \item \textbf{Gas depletion or fabric timeout}: it indicates that a transaction call in the faulty contract failed through a \emph{rollback} of the affected smart contract call due to gas depletion or fabric timeout,while it did not happen in the reference run. Gas depletion and Fabric timeout are respectively occur by EVM runtime implementation and error detection mechanisms in Hyperledger Fabric protection mechanisms. 

    \item \textbf{Reliability failure}: the result is a boolean and indicates whether there was at least one transaction in the faulty contract with a different result or return value than the reference contract. \emph{Reliability Failure} indicates (theoretically) client observable failures during output invariant checks.
  
    \item \textbf{Integrity failure}: the result is a boolean that indicates whether there was at least one successful transaction in the faulty contract with a different result or return value and a different read-write set (i.e., the faulty contract modified the state of the blockchain differently from the reference contract) than the reference contract. In this situation, the ledger integrity failure is also detectable by the client.
  
    \item \textbf{Latent integrity error}: the result is a boolean that indicates whether there was at least one successful transaction in the faulty contract with the same result or return value but a different read-write set than the reference contract. In this situation, the ledger integrity error stays hidden and cannot be directly detected by the client as the client receives the expected result or return value. 
    
    \item \textbf{Ineffective}: the result is a boolean that indicates whether the result, return value, and read-write set of all transactions calls of the faulty contract are same as the reference contract. In fact, faulty contracts passing even output invariant checks constitute cases where the injected fault remains \textbf{undetected} -- i.e., no fault-indicative error state is detected. These undetected faults are either \emph{ineffective}, or, on the contrary, constitute \emph{latent integrity errors}. 
    
\end{itemize}

The error/failure model, including the effects of smart contracts faults, and its characteristics are summarized in Table \ref{tab:failureModel}. In this table, to simplify, the effects of fault are categorized into three groups: i) Transaction not Concluded; ii) Incorrect return value or transaction result; iii) Incorrect ledger state. when a \emph{Abort failure} occurs, for any reason \cite{errorCodes}, neither a value or transaction result is returned to the client nor any changes is made in the state of ledger. Simply transaction is not concluded and some sort of exception or error message is given to the client allowing to detect the failure. When a \emph{Gas depletion or fabric timeout} occurs, same as \emph{Abort failure}, no value is returned and no changes to the ledger state is made. In this case, the fault causes the spent of more resources in terms of time and gas. Thus, the transaction is not concluded due to the lack of gas or system timeout. With \emph{Reliability failure}, the transaction is concluded but the return value or transaction result is incorrect, although the state of the ledger is still correct. In contrast, when an \emph{integrity failure} occurs, in addition to incorrect return value or transaction result, the integrity of the ledger state is violated as well. Either the lack of valid result (in the case of \emph{Abort failure} and \emph{Gas depletion or fabric timeout}) or incorrect result (in the case of \emph{Reliability failure} and \emph{Integrity Failure}) allows the client to detect the above issues. While in the case of \emph{Latent integrity error}, the integrity of the ledger state is violated but, since the return value to the client is correct, it remains undetected by the client. A fault is \emph{ineffective} when the faulty contract behaves exactly same as the reference contract.

\begin{table}[h]
    \centering
    \caption{Error/failure model and its characteristics.}
    \label{tab:failureModel}
    \begin{tabular}{c|c|c|c}
         \begin{tabular}{@{}c@{}} \\ \textbf{Fault Effect} \end{tabular} &
         \begin{tabular}{@{}c@{}}\textbf{Transaction} \\ \textbf{not Concluded} \end{tabular} &
         \begin{tabular}{@{}c@{}}\textbf{Incorrect Return Value} \\ \textbf{or Transaction Result} \end{tabular} &
         \begin{tabular}{@{}c@{}}\textbf{Incorrect} \\ \textbf{Ledger State} \end{tabular} \\
         \hline
         \hline
         Abort failures & \checkmark &  & \\
         \hline
         Gas depletion/Fabric timeout & \checkmark &  & \\
         \hline
         Reliability failures &  & \checkmark  & \\
         \hline
         Integrity failures &  & \checkmark & \checkmark \\
         \hline
         Latent integrity error &  &  & \checkmark \\
         \hline
         Ineffective &  &  \\
         \hline
    \end{tabular}
\end{table}

%% file: 6-results.tex
\section{Results}
\label{sec:results}

This section presents the results obtained during our experimental evaluation. It first overviews the effects caused by the different injected faults. Then, the results of evaluation of the effectiveness of the formal verification and other protection mechanisms are presented. At the end, the preliminary performance evaluation results are presented.

\subsection{Fault Effects Overview}

\begin{figure}[h]
\centering
 \includegraphics[width=0.90\linewidth]{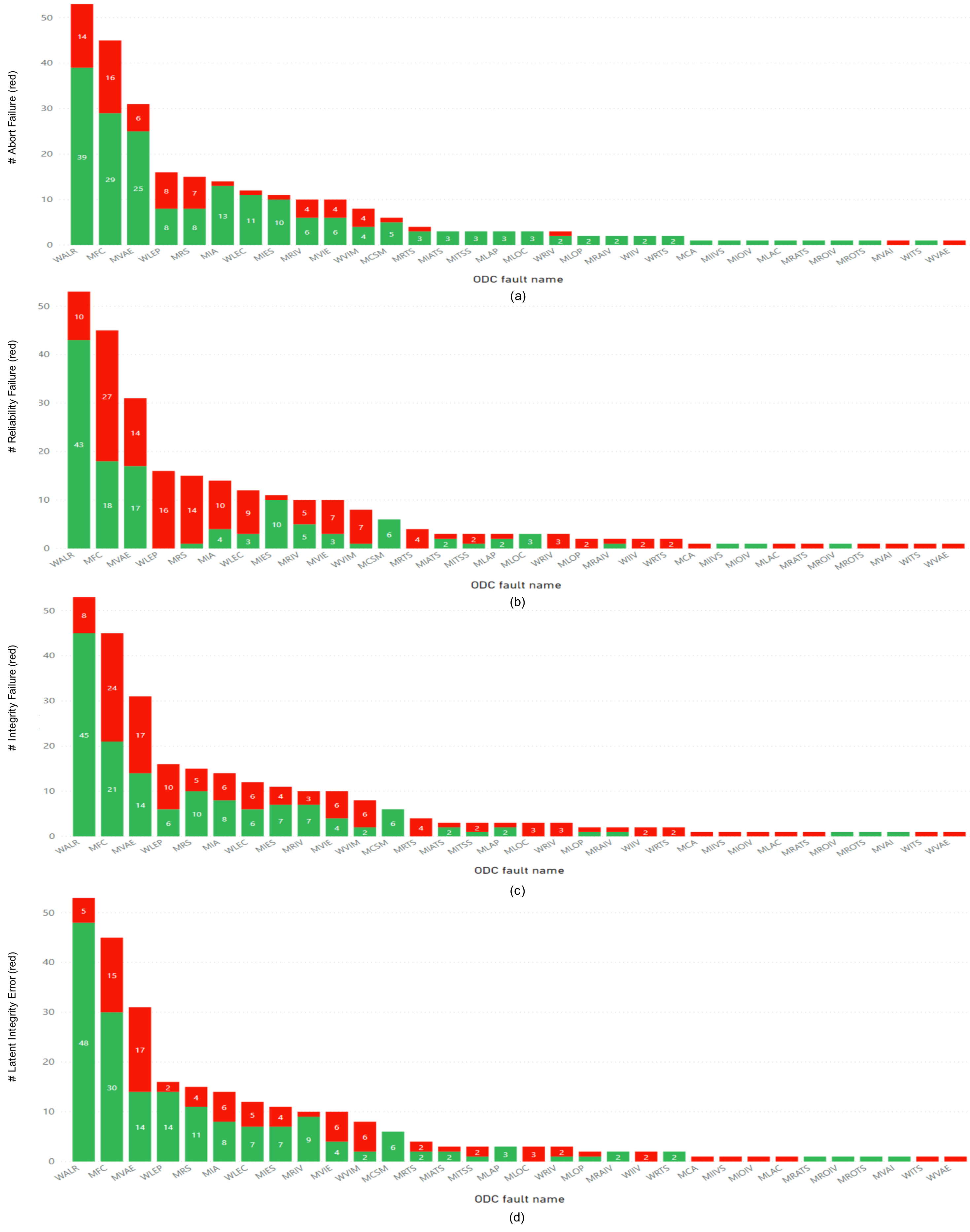}
 \caption{Faults impact.}
 \label{fig:faultsImpact}
\end{figure}

We begin by presenting the distribution of injected faults and their impact in terms of the error/failure model, previously defined in Section \ref{subsec:result-analysis}, and with respect to the the base contracts only, as they are the most realistic versions. Fig. \ref{fig:faultsImpact} shows the distribution/impact of the faults, with the different faults being listed from the most frequent to the least frequent ones in the x-axis. For instance, \emph{Wrong Algorithm (WALR)}, which is the most frequent fault, is present in many of the faulty contracts and \emph{Wrong Arithmetic Expression Used in Assignment (WVAE)} is present in just a few faulty contracts. The frequency of the faults presented in Fig. \ref{fig:faultsImpact} does not directly imply the frequency of the mistakes that developers usually make, but relates to the number of possible places in the typical contracts' code, in which each type of fault may appear. The faults that are not present in the figures did not appear in any base contract. The very first observation is that, in general, the generic faults are more frequent than smart-contract specific faults.

Each of the four plots in Fig. \ref{fig:faultsImpact}, shows the number of times (i.e., the number of faulty contracts with a specific fault) each individual fault caused a specific problem regarding the error/failure model. For instance, Fig. \ref{fig:faultsImpact} (a) shows that from the 53 base faulty contracts with the \emph{Wrong Algorithm (WALR)} fault, \emph{abort failure} occurred in at least one of the transactions executed against 14 faulty contracts (red part in the figures), and as shown in Fig. \ref{fig:faultsImpact} (b), \emph{reliability failures} occurred in at least one of the transactions of 10 faulty contracts, and so on. Note that each fault may cause more than one failure and error in the same contract. 
In total, 71 injected faults caused \emph{abort failures}; 145 injected faults caused \emph{reliability failures}, 122 injected faults caused \emph{integrity failures}; and 90 injected faults caused \emph{latent integrity errors} in at least one of the transactions executed against the faulty contracts. Thus, reliability failure is the most frequent failure and abort failure is the least frequent one.

As we can see in Fig. \ref{fig:faultsImpact} \emph{Missing function call (MFC)} is highly frequent and most errors and failures, specially \emph{reliability failures} and \emph{integrity failures}, are caused due to this fault (a total of 82 failures/errors). \emph{Latent integrity error} that is the most serious issue is mostly caused by \emph{Missing variable assignment using an expression (MVAE)} and \emph{Missing function call (MFC)}. On the other hand, \emph{Missing require OR subexpression on input variables (MROIV)} is the only one that did not cause any failure or error; it was not a frequent fault in base contracts and when happened did not cause any failure or error. \emph{Wrong arithmetic expression used in assignment (WVAE)} is not frequent either but, when occurred, caused all kind of failures and errors.

\emph{Wrong logical expression in parameters of function call (WLEP)} is present in 16 faulty contracts and in all cases triggered a reliability failure. 
\emph{Missing return statement (MRS)}, \emph{Missing if construct around statement (MIA)}, and \emph{Wrong logical expression used as branch condition (WLEC)} also mostly caused reliability failures.

Among smart contract specific faults, \emph{Missing require on input variables (MRIV)} and \emph{Missing call to SafeMath (MCSM)} are more frequent faults. \emph{Missing require on transaction sender (MRTS)}, \emph{Wrong logical expression in require over input variables (WRIV)}, \emph{Wrong logical expression in if construct over input variable (WIIV)}, and \emph{Wrong logical expression in require over transaction sender (WRTS)}, although not being very frequent, caused reliability and integrity failures in all faulty contracts in which they were injected. 

\subsection{Base Contracts Results}

We visualize the sequence of successively escaping faulty contracts using Sankey-like diagrams. In these diagrams, blue rectangles represent protection mechanisms and ellipses are classification of system output, referring to cases where a given contract was indeed labelled as faulty at a given phase. The order of steps is quite typical, representing a suggested development workflow and the protection logic of EVM, while the thickness of the lines shows the ratio of contracts passing a check. First, we analyze the detection results for all faulty instances of the base contracts, measured as described earlier (refer to Section \ref{subsec:protectionLevel} and Section \ref{subsec:result-analysis}). 

\begin{figure}[h]
\centering
 \includegraphics[width=0.7\linewidth]{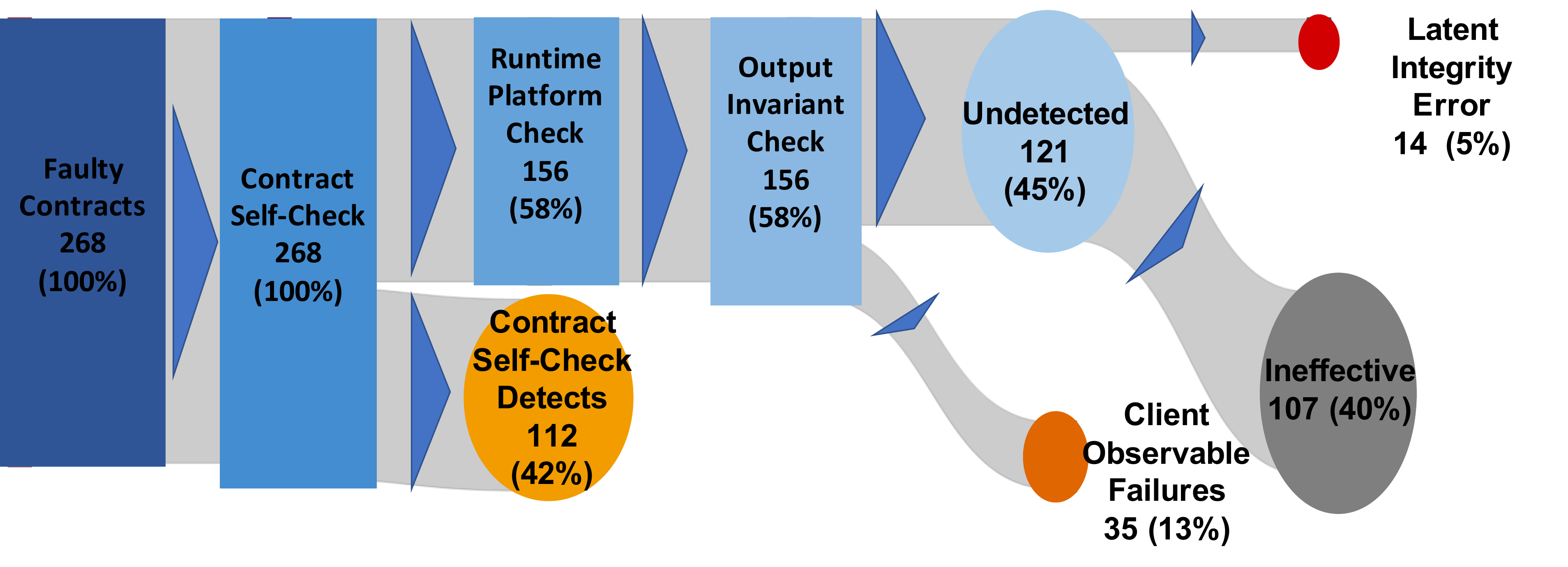}
 \caption{Fault detection in \textit{base contracts} without formal verification step.}
 \label{fig:sankey-noverif-baseline}
\end{figure}

\begin{figure}[h]
\centering
 \includegraphics[width=0.7\linewidth]{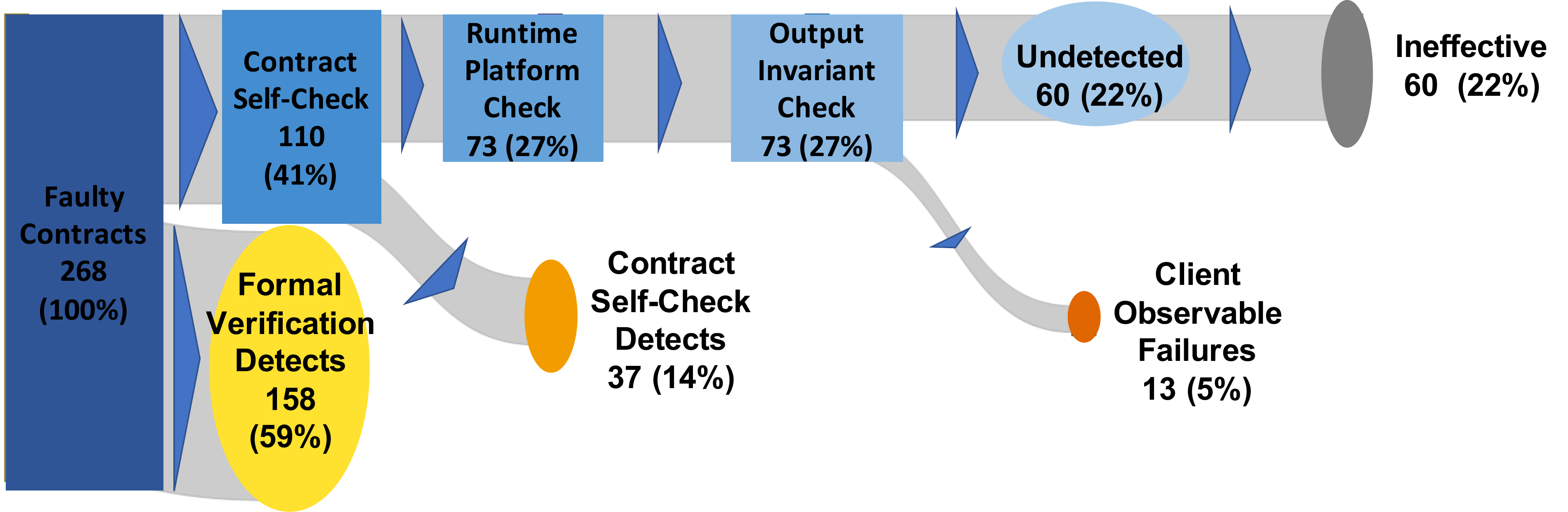}
 \caption{Fault detection in \textit{base contracts} with formal verification step.}
\label{fig:sankey-verif-baseline}
\end{figure}

Fig. \ref{fig:sankey-noverif-baseline} shows the number of faulty base contracts eluding the different detection phases, but assuming that no verification is performed before deployment, while Fig. \ref{fig:sankey-verif-baseline} shows the number of faulty contracts escaping detection at the different phases, which include verification as part of the workflow. The number of generated faulty base contracts is 268, which is visible at the left-hand side in Fig. \ref{fig:sankey-noverif-baseline} and Fig. \ref{fig:sankey-verif-baseline}.

When verification is part of the workflow (refer to Fig. \ref{fig:sankey-verif-baseline}), all faulty contracts are first subject to formal verification, which detects $159$ faults out of $268$ (59\% of all faulty contracts).
The next step, contract-level self-check, expressed in the form of require and assert constructs, catches faulty contracts where a transaction terminated with an endorsement result different from the reference run. Contract-level self-check detects $37$ out of $110$ (which is 14\% of all 268 faulty contracts) faults that remained undetected by formal verification. 

The next step involves platform-level runtime checks. These checks refer to built-in EVM and Hyperledger error detection mechanisms (e.g., detection of timeout, out-of-bound array addressing, etc.). Violating such checks corresponds to a behavior which in a non-FT environment could result in severe failures, including system crashes. In the current experiments no faulty contracts were detected by these checks. 

The last level of protection mechanisms is the client-side validation of results (i.e., output invariant check in Fig. \ref{fig:sankey-noverif-baseline} and Fig. \ref{fig:sankey-verif-baseline}). In total, $13$  out of $73$ (5\% of all faulty contracts) faults that remained undetected by previous verification and checks were detected in this final step.  In general, the remaining faults which are undetectable by this step either cause \textbf{Latent Integrity Errors} or are \textbf{Ineffective}. The first case represents the most critical class of failures referring to a contract which may produce corrupted, uncovered ledger content, while mutants having only ineffective faults cannot be distinguished from the reference (i.e., their injection, in combination with the workload, did not produce any change in the contract behavior). In the case of baseline mutants with formal verification (refer to Fig. \ref{fig:sankey-verif-baseline}) all $60$ remaining faults (22\% of all mutants) are ineffective.

The results, presented in Fig. \ref{fig:sankey-noverif-baseline} and Fig. \ref{fig:sankey-verif-baseline}), suggest that for contracts with ``reasonable'' protections (i.e., the \textit{base contracts}, which apply simple defensive programming techniques), verification and runtime checks are complementary techniques for detecting faulty contracts with potentially effective faults. On one hand, verification applied on the faulty contracts caught only 59\% of the faults. On the other hand, verification does filter out those 14 faulty contracts (5\% of all mutants) which led to latent integrity errors (see Fig. \ref{fig:sankey-noverif-baseline}).

Faults that are not ineffective, but pass verification and self-checks are typically related to liveness or insufficient specification. In the former case (liveness), faults cause to revert transactions that should not revert. Formal verification currently targets safety property (and not liveness) and does not detect such issues. Furthermore, self-checks have no chance to catch these issues as the transactions are reverted before the checks. In the latter case, specification was not sufficient enough, e.g., for the state machine contract we did not specify the expected behavior for all the (roughly 20) state variables.

One particularly interesting case escaping both verification and self-checks was the missing auto increment for loop variables. This caused an infinite loop which was forcefully terminated by the ledger. Catching this error in Solc-verify requires reasoning about termination (which is complex and not supported, at the time of writing) and catching them by self-checks would require complex assertions on the progress of loops.

Note that latent integrity errors are arguably the worst failure mode for distributed ledgers. As ledgers are essentially built over distributed databases, their generally expected operational model explicitly rules out transaction rollback. As a consequence, to repair data corruption, the only viable options are governance processes (by blockchain peer consortium) and embedding compensatory mechanisms into the on-chain workflows.

We deem the significant ratio of ineffective faults normal. Solidity smart contract testing has to deal with the extremely large Ethereum address space; until such time that symbolic automated test pattern generation becomes available for Solidity and/or the EVM, the only viable option is to, as we did, manually define the workload to be used. Evaluating test coverage is part of further research.

\begin{figure}[h]
\centering
 \includegraphics[width=1.0\linewidth]{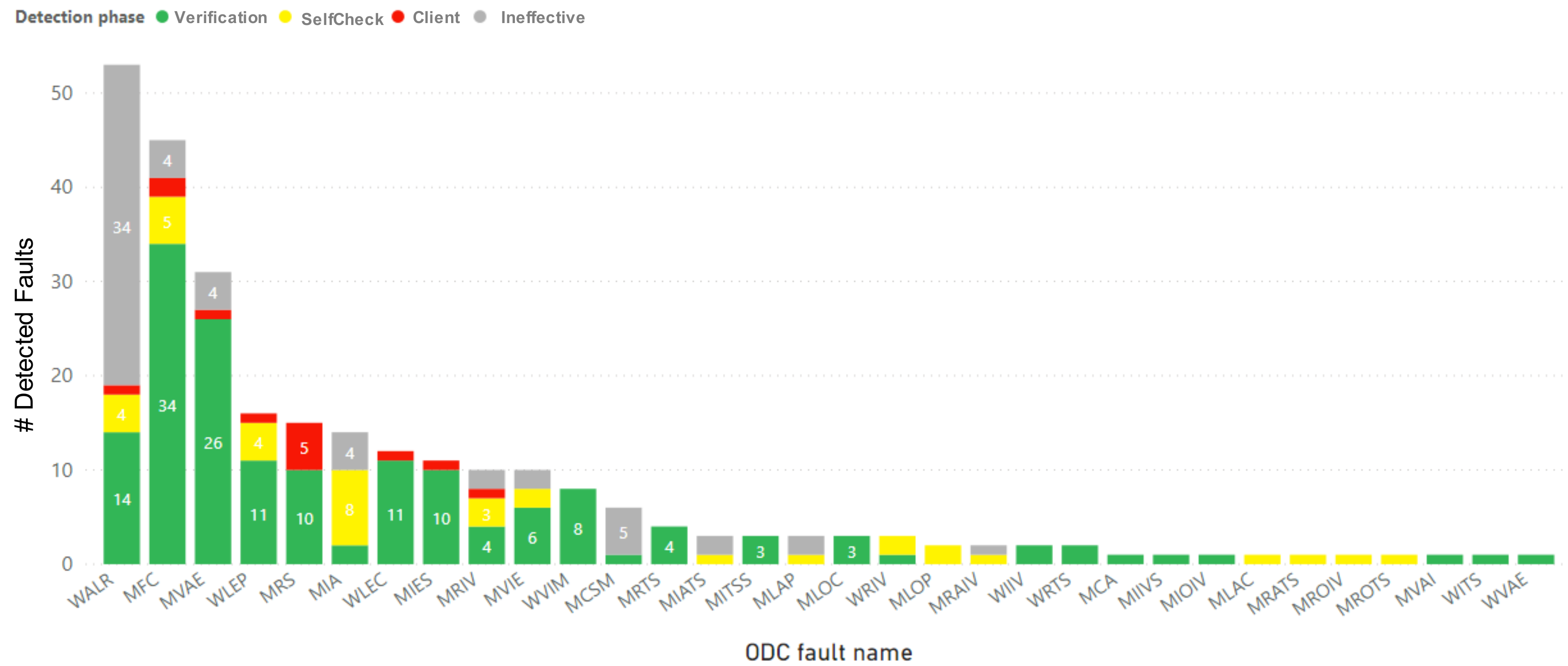}
 \caption{Effectiveness of formal verification and protection mechanisms on each fault in the base contracts.}
\label{fig:baseFaultDetection}
\end{figure}

Fig. \ref{fig:baseFaultDetection} presents the results of the effectiveness of formal verification and fault detection mechanisms in the base faulty contracts, by showing in which detection phase the injected faults are detected.

The results show that \emph{Missing function call (MFC)} and \emph{Missing variable assignment using an expression (MVAE)} that cause more serious failures and errors are mostly detected by formal verification. 
Similarly, \emph{Wrong logical expression in parameters of function call (WLEP)}, \emph{Missing return statement (MRS)}, and \emph{Wrong logical expression used as branch condition (WLEC)} (that cause reliability failures in most cases) are also mostly detected by formal verification. However, formal verification does not seem to be very effective on cases like
\emph{Missing if construct around statement (MIA)}. On the other side, contract self-checks appear to be effective in detecting this fault.  

Among all faults \emph{Missing return statement (MRS)} is the one that, after applying formal verification and protection mechanisms, causes more reliability/integrity failures, but is detectable by the client (red bar in the figure). Among other problematic smart contract specific faults, \emph{Missing require on transaction sender (MRTS)}, \emph{Wrong logical expression in if construct over input variable (WIIV)}, and \emph{Wrong logical expression in require over transaction sender (WRTS)} are fully detected by formal verification. However, the same does not happen with \emph{Missing call to SafeMath (MCSM)} and \emph{Wrong logical expression in require over input variables (WRIV)}. 

The final general observation is that, after applying verification and protection over the base contracts, only a few faulty contracts (13 out of 268 , showed in red) cause reliability or integrity failures and all these cases are detectable by the client. Obviously, the difficulty lies in applying the correct mechanisms at the correct time, and fault injection may help in this regard by identifying cases for which developers must pay special attention.

\subsection{Protected Contracts Results}
\begin{figure}[h]
\centering
 \includegraphics[width=0.7\linewidth]{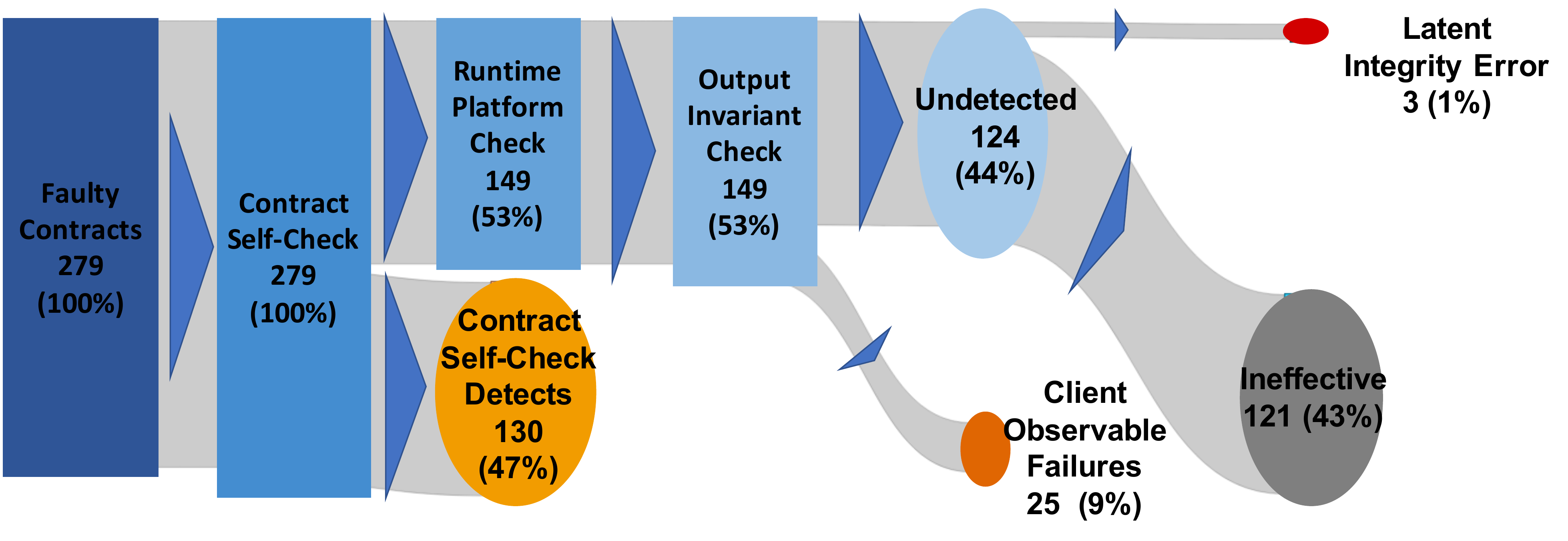}
 \caption{Fault detection in \textit{protected contracts} without formal verification.}
 \label{fig:sankey-noverif-protected}
\end{figure}

\begin{figure}[h]
\centering
 \includegraphics[width=0.7\linewidth]{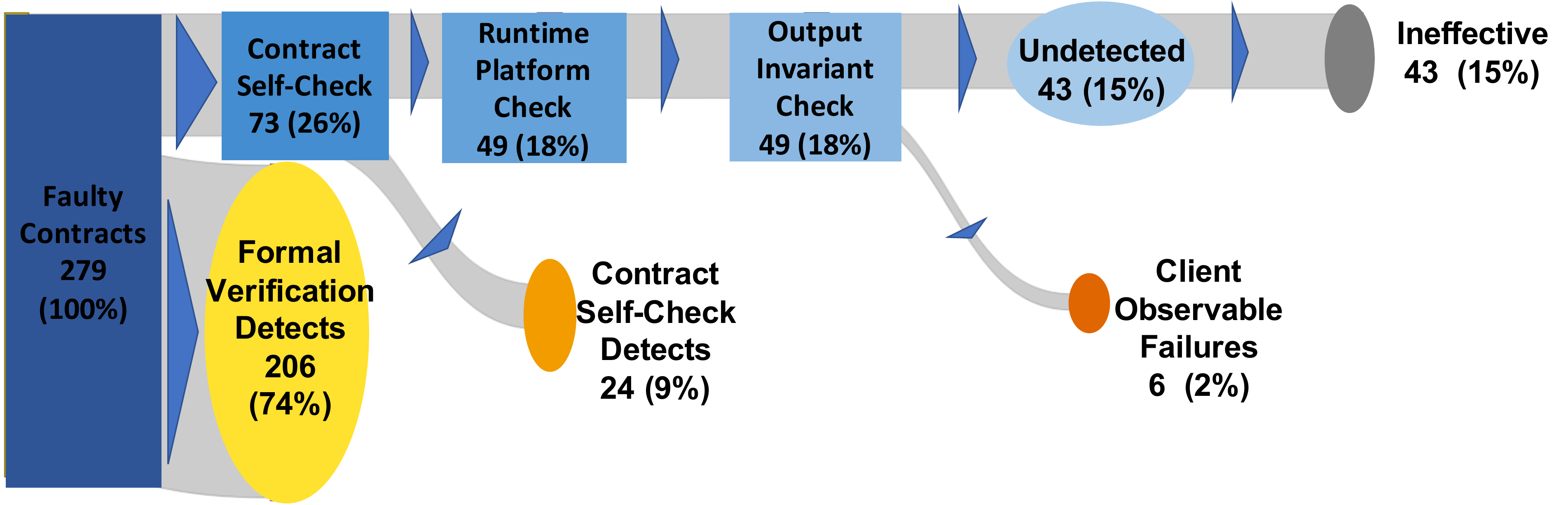}
 \caption{Fault detection in \textit{protected contracts} with formal verification.}
 \label{fig:sankey-verif-protected}
\end{figure}

Fig. \ref{fig:sankey-noverif-protected} and Fig. \ref{fig:sankey-verif-protected} show the number of faulty \textit{protected contracts} escaping detection at the different phases (a total of 279 protected faulty contracts). Fig. \ref{fig:sankey-noverif-protected} refers to the case where verification is used, whereas \ref{fig:sankey-verif-protected} refers to the case where verification is not used. The different number of faulty contracts is due to slight differences in source code. While \texttt{assert} statements are not subjected to faults, their supporting modifications are. 

In comparison to the \textit{base contracts} case presented in the previous section, the following three main findings can be formulated:

\begin{itemize}[-]
    \item Assuming no verification, \emph{even ``cheap'' and simple post-condition checking (on-chain and client) significantly reduce the ratio of latent integrity errors} and the significance of client-side checks is slightly reduced.
    \item Verification becomes significantly more effective; it can use the protective statements as ``additional specification''. However, we also experienced false positives for the Token contract, due to the verifier not being able to prove assertions in a loop.
    \item Combined verification and runtime checks eliminate latent integrity errors in both cases, and could also reduce the number of client observable faults.
\end{itemize}

\begin{figure}[h]
\centering
 \includegraphics[width=1.0\linewidth]{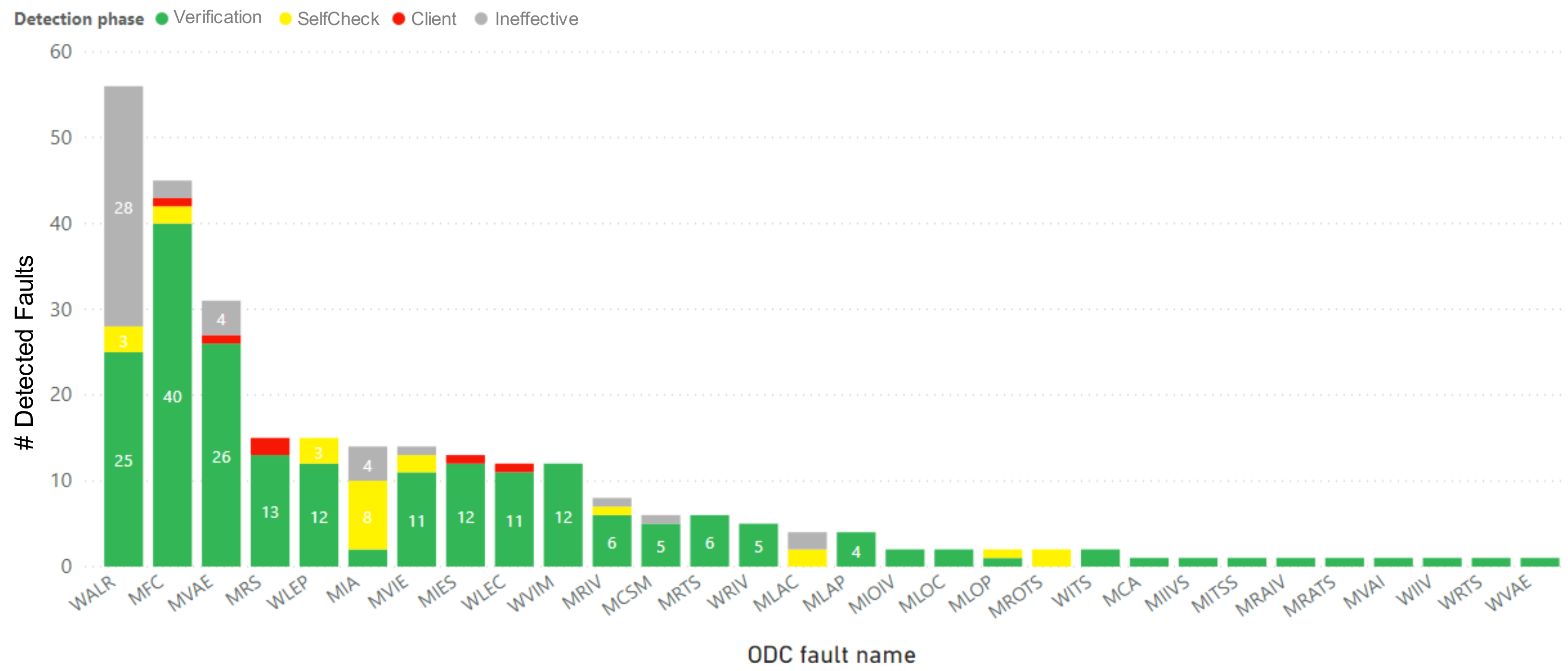}
 \caption{Effectiveness of formal verification and protection mechanisms on each fault in the protected contracts.}
\label{fig:protectedFaultDetection}
\end{figure}

Fig. \ref{fig:protectedFaultDetection} presents the results of the effectiveness of formal verification and fault detection mechanisms for the protected contracts. The types of faults that could be injected into the protected contracts are exactly the same as the faults that are injected into the base contracts, although their order in terms of frequency is slightly different due to the specificities of the additional protection mechanisms present in the code.

After analysing Fig. \ref{fig:protectedFaultDetection} and Fig. \ref{fig:baseFaultDetection}, we observed a very similar pattern regarding the effectiveness of the formal verification and protection mechanisms across these two groups of contracts. However, after applying verification and protection to the protected contracts, a lower number of faulty contracts (6 out of 279 , showed in red) caused reliability or integrity failures (detectable by the client), when compared to the base contracts.  

\subsection{Stripped Contracts Results}

\begin{figure}[h]
\centering
 \includegraphics[width=0.7\linewidth]{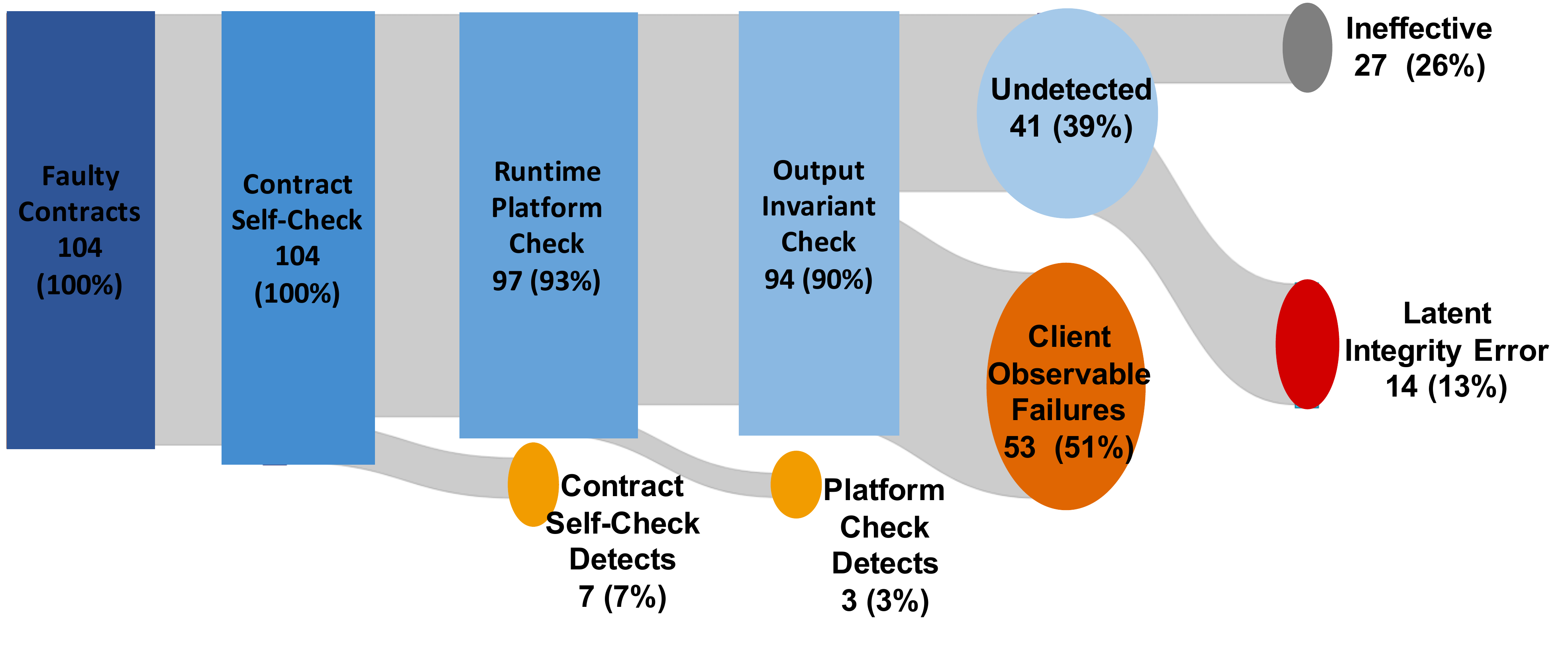}
 \caption{Fault detection in \textit{stripped contracts} without formal verification.}
 \label{fig:sankey-noverif-stripped}
\end{figure}

\begin{figure}[h]
\centering
 \includegraphics[width=0.7\linewidth]{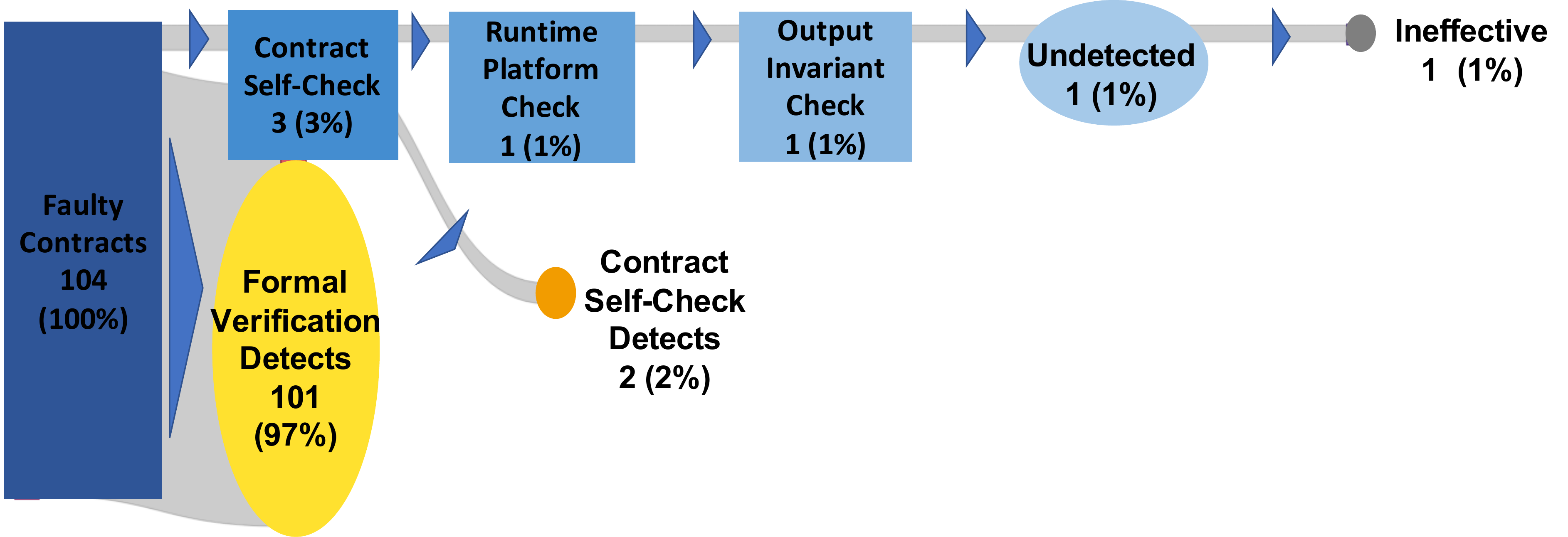}
 \caption{Fault detection in \textit{stripped contracts} with formal verification.}
 \label{fig:sankey-verif-stripped}
\end{figure}

Fig. \ref{fig:sankey-noverif-stripped} and Fig. \ref{fig:sankey-verif-stripped} show the number of faulty \textit{stripped contracts} (a total of 104 stripped faulty contracts) escaping detection at the different phases respectively in the cases without and with formal verification. As shown, for stripped contracts without verification, more than half of the mutants results in client observable failures, as contract self-check and some built-in EVM mechanisms (like balance check) can catch only a few of the injected faults (10\% in total). From the remaining faulty contracts, 13\% of them result in latent integrity error, which is quite high. 
Formal Verification, on the other hand, catches 97\% of the faulty contracts and only one (ineffective) fault passes all protection mechanisms. 

The results, on one hand, confirmed our presupposition that eliminating even the basic runtime precondition checks leads to unacceptable levels of latent integrity errors. On the other hand, as we used the same precondition specifications for all contracts in a family, verification could trivially find counterexamples in all cases due to the missing \texttt{require} statements.

\begin{figure}[h]
\centering
 \includegraphics[width=1.0\linewidth]{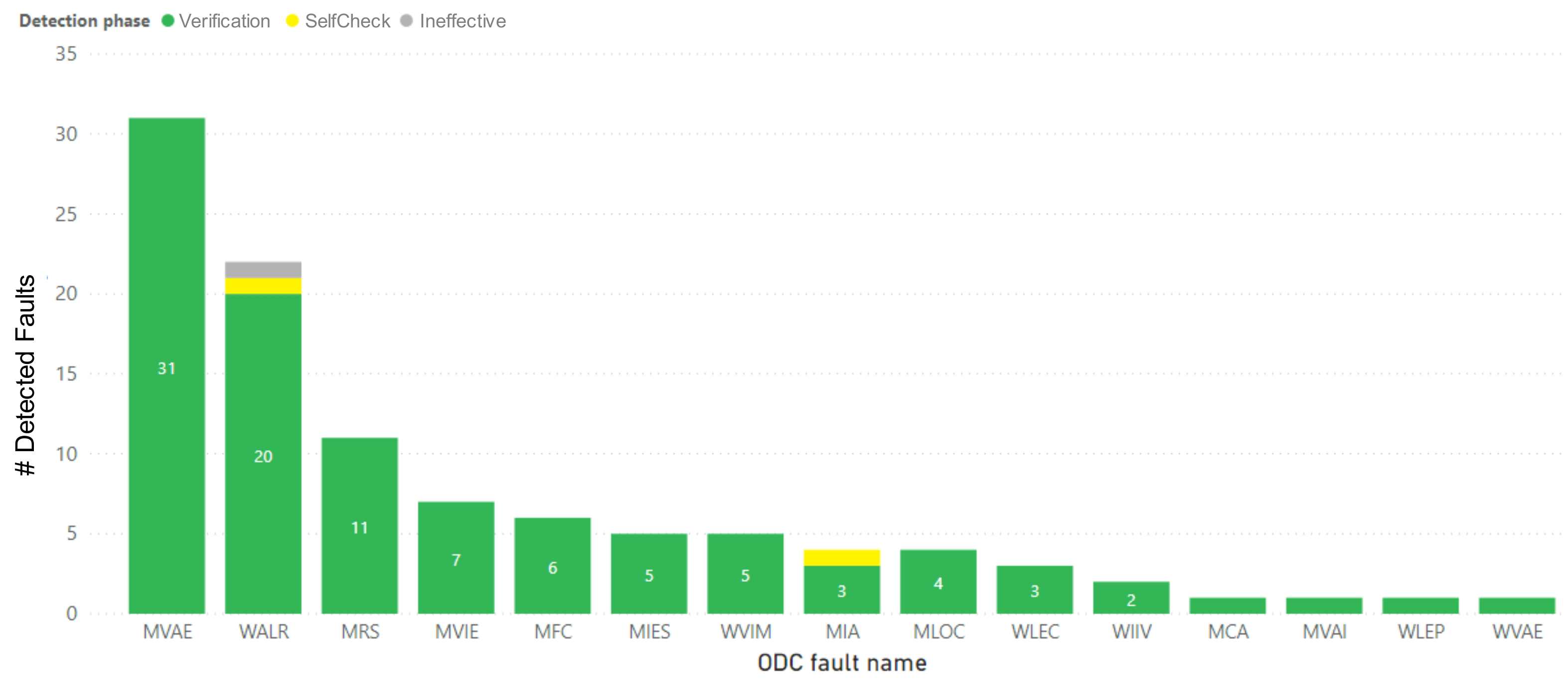}
 \caption{Effectiveness of formal verification and protection mechanisms on each fault in the stripped contracts.}
\label{fig:strippedFaultDetection}
\end{figure}

The above observation is confirmed in Fig. \ref{fig:strippedFaultDetection}, which presents the results of the effectiveness of formal verification and fault detection mechanisms for the stripped contracts. As we can see, a smaller number (i.e., when compared to the base or protected contracts) of types of faults could be injected into stripped contracts. The main aspect visible is that almost all cases are detected by formal verification.

\subsection{Verification Effectiveness}

During our experiments, we observed latent integrity errors, but only when formal verification was not applied (i.e., $3$ latent integrity errors in the protected contracts and $14$ in the base contracts). We investigated all cases and came to the conclusion that all latent integrity errors were caused by missing constructs, for example missing variable initializations or missing subexpressions in the conditions of \texttt{if} statements. 
While the errors could have also been avoided with more protection, we observed that writing formal specification was, in some cases, more efficient than writing extra protection. For instance, instead of defining invariants at function level, contract-level constructs can also be applied \cite{solcverify}.

To evaluate the effectiveness of verification and impact of additional protections, we summarized the results by comparing the \emph{recall} values of base and protected variants. \emph{recall} is calculated by dividing the number of detected faulty contracts (true positives) over the total number of faulty contracts (the sum of true positives and false negatives). Table \ref{tab:recall-stats} presents the ``verification recall'', evaluated for different groups of faulty contracts with common error/failure effects (e.g., faulty contracts causing ledger integrity violation, faulty contracts causing reliability failures).

\begin{table}[h]
    \centering
    \caption{Verification recall for faulty contracts with different error/failure effects.}
    \label{tab:recall-stats}
    \begin{tabular}{c|c|c}
         \begin{tabular}{@{}c@{}}\textbf{Faulty contract set by} \\ \textbf{error/failure effect} \end{tabular} &
         \begin{tabular}{@{}c@{}}\textbf{Verification recalls for} \\ \textbf{base variants} \end{tabular} &
         \begin{tabular}{@{}c@{}}\textbf{Verification recalls for} \\ \textbf{protected variants} \end{tabular} \\
         \hline
         \hline
         Abort failures & 69\% & 79\% \\
         \hline
         Gas depletion/Fabric timeout & 100\% & 100\% \\
         \hline
         Reliability failures & 67\% & 81\% \\
         \hline
         Integrity failures & 78\% & 84\% \\
         \hline
         Latent integrity error & 83\% & 76\% \\
         \hline
         Ineffective & 69\% & 81\% \\
         \hline
         
         \hline
         All effective faults & 69\% & 81\% \\
    \end{tabular}
\end{table}

The results show that runtime protections added into the protected contracts, acting as additional specification, improve the effectiveness of formal verification (i.e., recall increases) in almost all cases. Hidden side effects is the exception. The reason for the lower recall in this case is that the ``return value introduction'' protective measure changes the ``hidden side effects'' baseline failure category of a number of mutants through increased observability (as also expressed on ~\ref{fig:sankey-verif-baseline}). Ratio-wise, the remaining cases of the protected mutants (no new ones introduced) become harder to catch with verification.

Due to restrictive default Burrow settings in the measurement setup, Fabric timeouts (1 second) were never engaged; thus, the ``EVM gas depletion or Fabric timeout'' cases actually only ever exercised the EVM mechanism that limits the number of computational steps. While verification caught all such cases (e.g., infinite loops), we performed a dedicated experiment to also measure active Fabric-level protections, which is presented in the following section (Section \ref{subsec:quantitative-analysis}).

\subsection{Preliminary Assessment of Blockchain Performance}
\label{subsec:quantitative-analysis}

As a complement to the experimental evaluation, we carried out a set of experiments, of anecdotal nature, aiming at understanding performance issues related with the execution of faulty contracts. \textit{Performance evaluation} allows to understand if there are significant changes in execution time of faulty smart contracts, when compared to their reference counterparts. Protection mechanisms defined in the blockchain system (e.g., definition of gas limit) do not allow smart contracts to execute indefinitely, hence, performance problems can eventually be transformed to correctness problems (i.e., failed transactions). Additionally, and especially in permissioned systems, the different resource usage profiles of faulty contracts may have an impact on overall blockchain system performance and availability through resource contention. In this work, performance evaluation is mainly concerned about the execution time of transactions. 

We performed our experiments using the \texttt{Token} \textit{base contract}, with an infinite loop fault in its \texttt{batchTransfer} function and no fault in its \texttt{transfer} function (standard token management functions). The test sequence calls i) the fault-free function 400 times (warm-up fault-free period) to be used as reference and baseline behavior of the system for the comparison; ii) the faulty function 200 times; and iii) the fault-free function 400 times again to be compared with reference runs in order to understand how the faulty runs impacted on the execution time of the fault-free function. The calls were all data-independent. Burrow was reconfigured to enable execution to hit (and possibly far exceed) the Fabric timeout, but the measurement environment otherwise remained unchanged. The number 400 and 200 are arbitrarily chosen but in a way that let us to observe the behavior of the system long enough.

\begin{figure}[h]
\centering
 \includegraphics[width=0.6\linewidth]{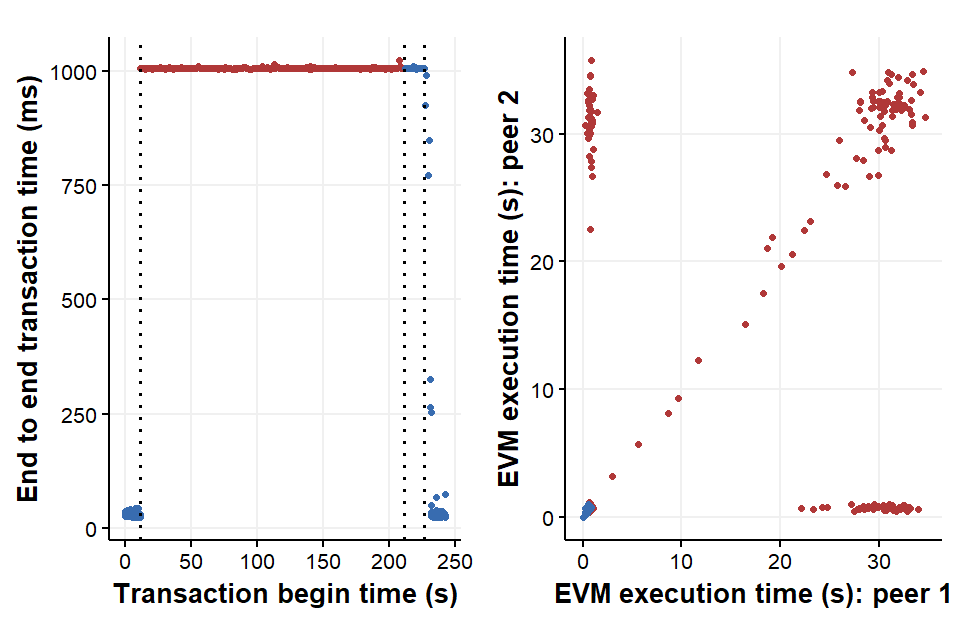}
 \caption{End-to-end transaction (attempt) times and Burrow EVM execution times in the timeout campaign (red: faulty, blue: fault-free function).}
 \label{fig:timeouts}
\end{figure}

Fig. \ref{fig:timeouts} shows the client-perceived transaction times as well as the call execution time measured \emph{inside the EVM} through additional instrumentation. The colors distinguish the two functions: red for the faulty function and blue for fault-free function.
As shown in the figure, upon switching to the faulty function, end to end transaction time increases to around 1 s (the Fabric timeout) and transactions fail; however, when the workload switches back to the fault-free function, for approximately 14 seconds it remains at 1 second before regaining the nominal value (and transactions begin to succeed). At the same time, EVM executions are not \emph{actually} terminated in the containers, leading to increasing resource contention.
 
Evidently, this is a resource management bug in the blockchain environment (CPU-overloading runoff executions are not terminated and garbage-collected in a timely manner). The significance of this phenomenon in our context is that it demonstrates that, in general, we \emph{cannot} blindly rule out the possibility of smart contract faults having an extra-functional impact on the execution platform and other smart contracts; especially in ``business'' blockchains that run more complex and comparatively less security and resilience vetted platform software than the public ones.

%% file: 8-conclusion.tex

\section{Threats to Validity}
\label{sec:threats}

In this section, we present the main threats to the validity of this work and discuss mitigation strategies. We start by mentioning that \textit{the number and types of smart contracts used in this work are limited}. We are also aware that libraries for domain specific smart contract development, such as OpenZeppelin or Hyperledger Grid are rapidly maturing. Smart contract development is also increasingly utilizing domain-specific languages and generative/interpretative techniques, as available for, e.g., DAML and Hyperledger Concerto. We plan to apply the techniques introduced in this paper for these technologies and over a large set of smart contracts in further research.
    
\textit{Our mutant generation approach can introduce some mutations that lead to code that is strictly behaviorally equivalent to the reference} (e.g., swapping two subsequent, data-independent, non-control statements or removing a local overflow check, which is not necessary due to additional global checks). Although it is not affecting the assessment results, filtering these out during mutation generation is ongoing work to avoid duplicate executions. 
    
\textit{The defensive mechanisms studies in this work are limited to formal verification and some additional solidity- and platform-level protection mechanisms}, which are the most popular mechanisms. However, further research will target investigating the effect of introducing further defensive techniques as business-logic derived runtime assertion checks, N-version programming, or, indeed, much of the wide range of fault-tolerant software patterns \cite{hanmer2013patterns}. These can prove especially important in enterprise blockchain systems, where sizeable, complex smart contracts and critical applications are rapidly appearing.
    
\textit{The manual definition of the fault model} may involve some error due to the human intervention in the process. Thus, some faults may be misclassified, which may affect their overall frequency. To mitigate this threat, the classification was verified by a second researcher, allowing for stronger confidence in the fault model.

\textit{Executing one transaction at a time may not be a representative scenario}, especially when considering performance metrics. Still we consider it to be a starting point to execute experiments that may impact performance and future work may consider scenarios involving the execution of multiple transactions at a time.

\textit{The performance analysis of Blockchain system is limited to the execution time} of one faulty transaction compared to its reference run, which is quite relevant. Nevertheless, this study will be extended in future to include the whole fault model considering other performance attributes. 

\section{Conclusion} 
\label{sec:conclusion}

In this paper we presented an approach for assessing key dependability properties of blockchain systems, in presence of faulty smart contracts. The approach is based on the injection of a set of software faults that is composed by general cases of software faults and blockchain-specific cases. We execute runtime tests over the modified faulty contracts to observe divergence from reliable (or correct) behavior and ledger state integrity. We also evaluate the effectiveness of formal verification and contract-level defense and protection mechanisms against the injected software faults.

We evaluated a small but -- considering the nature of smart contracts -- representative set of contracts and found that the combined use of formal verification and protection mechanisms is a powerful way of reducing the number of client observable and residual faults. Still, these techniques are not sufficient to prevent the occurrence of all kinds of faults, with a few cases escaping these verification techniques. We summarize the main findings and observations in the next paragraphs.

Modern blockchain frameworks support the application developer by offering built-in services for the most complex tasks, especially by managing data distribution and synchronization, consensus, and ordering of transactions. Despite this, \textbf{current built-in verification tools and services lack in adequate support to assure the dependability of such complex systems.} The principle of replicated synchronized ledgers and prior agreement on contracts and their execution by consensus implements an n-out-m modular redundancy scheme providing system-level guarantees for Byzantine or random errors.
However, \textbf{as software faults in accepted smart contracts corrupt all nodes in a common mode way, replication and consensus are ineffective and they remain uncovered.} 

Solidity and its run-time environments offer limited mechanisms for self-checking and validation of user inputs in the form of assumptions and assertions. \textbf{Enrichment of the contract with assertions and requirements reveals a significant number of logical errors} by blockchain assertion checks. Further \textbf{platform level checks detect additional faults} before committing transactions (e.g., timeouts, account validation, balance validation in case of native tokens). \textbf{Formal verification can cover a broader subset of faults} and doing the checks at design time can reduce execution cost by omitting the already proven assertions. In our pilot experiments, latent errors were eliminated and number of faults that escape the detection and protection mechanisms in the blockchain system (resulting in client observable failures) was also significantly reduced.

As future work, we aim to focus on security aspects of blockchain systems. The fault model created and presented in this paper includes several faults that can be exploited by attackers leading to severe security issues (e.g., missing \emph{require} on transaction sender (MRTS) or wrong visibility (public) for private\/internal function (PVPF)). In order to properly assess the impact of these kinds of faults/security vulnerabilities, security penetration testing is a required technique which, to the best of our knowledge, has been scarcely explored in the blockchain domain. There is a need for effective tools specialized in security for blockchain systems and that are ready to handle the specifities of these systems. This need is increasingly important, especially considering the rapid adoption of blockchain systems and their new applications, but is especially difficult due to challenges like the definition of effective attackloads and also the evaluation of the impact of an attack, which must be performed on a distributed, complex system.